%% file: main.tex
\documentclass{article}

\usepackage[T1]{fontenc}
\usepackage{microtype}
\usepackage{graphicx}
\usepackage{subcaption}
\usepackage{booktabs}
\usepackage{hyperref}

\usepackage[preprint]{icml2026}

\usepackage{amsmath}
\usepackage{amssymb}
\usepackage{mathtools}
\usepackage{amsthm}
\usepackage[capitalize,noabbrev]{cleveref}
\usepackage{pgfplots}
\usepgfplotslibrary{groupplots}
\pgfplotsset{compat=1.18}
\usepackage{enumitem}

\theoremstyle{plain}
\newtheorem{theorem}{Theorem}
\newtheorem{proposition}{Proposition}
\newtheorem{lemma}{Lemma}
\newtheorem{corollary}{Corollary}
\theoremstyle{definition}
\newtheorem{definition}{Definition}
\newtheorem{assumption}{Assumption}
\theoremstyle{remark}

\newcommand{\R}{\mathbb{R}}
\newcommand{\E}{\mathbb{E}}
\newcommand{\Prob}{\Pr}
\newcommand{\AM}{\mathrm{AM}}
\newcommand{\GM}{\mathrm{GM}}
\newcommand{\HM}{\mathrm{HM}}
\newcommand{\tr}{\operatorname{tr}}
\newcommand{\diag}{\operatorname{diag}}
\newcommand{\supp}{\operatorname{supp}}
\newcommand{\Haar}{\mathrm{Haar}}
\newcommand{\cP}{\mathcal{P}}

\newcommand{\rot}{\mathrm{rot}}
\newcommand{\none}{\mathrm{none}}
\newcommand{\WS}{\mathcal{U}}  
\newcommand{\Cw}{C_{\mathrm w}} 
\newcommand{\dist}{\mathsf{D}}  
\newcommand{\hd}{\mathrm{h}}    
\newcommand{\Neg}{\mathrm{Neg}} 

\icmltitlerunning{How Much Does Random Rotation Help GPTQ?}

\hypersetup{pdfsubject={Research preprint: GPTQ, random rotations, and high-rate quantization}}

\begin{document}

\twocolumn[
  \icmltitle{How Much Does Random Rotation Help GPTQ?\\
    Sharp Worst-Case Rate Gaps in the High-Rate Regime}

  \input{authors}

  \icmlkeywords{post-training quantization, GPTQ, lattices, random rotation, free probability, rate-distortion}

  \vskip 0.3in
]

\printAffiliationsAndNotice{}

\begin{abstract}
GPTQ is Babai's nearest-plane algorithm on the lattice of the Cholesky factor of the layer's input second moment $\Sigma$. In the high-rate, entropy-coded regime it needs $\tfrac12\log_2(\AM/\GM)$ of its Cholesky pivots more bits than WaterSIC, which is within $0.255$ bit of the rate-distortion limit for every $\Sigma$; random rotations close most of this gap in practice, and Ordentlich and Polyanskiy asked for the worst-case gap between rotated GPTQ and WaterSIC. We show that after a Haar rotation the $k$-th pivot converges almost surely, uniformly in $k\ge\delta n$ and over spectra $\nu$ of bounded condition number, to $1/S_\nu(-k/n)$, a classical formula from the theory of Lyapunov exponents, and use it to answer the question. Over spectra with condition number $\kappa$ the worst rotated gap is $\tfrac12\log_2\kappa-\log_2\ln\kappa+\tfrac12-\tfrac1{\ln2}+o(1)$ bits, attained only by two-point spectra, just $\tfrac12\log_2\ln\kappa+O(1)$ bits below the worst case without rotation, so rotation cannot replace waterfilling in general. It nearly can on real layers: rotation cuts the cost of an $\eta$-fraction of outlier eigenvalues of size $\mu$ from about $\eta\mu$ to $\eta^2\mu/2$ to leading order, but cannot remove spectral spread on a logarithmic scale; for non-Gaussian weights it also adds a weight-side rate cost, which we characterize exactly under a standard high-rate assumption. On $96$ Hessians of Qwen3 and Llama-3.1 the rotated gap has median $0.04$ bit and maximum $0.36$ bit, against $2$--$5$ bits in the worst case for the same condition numbers, and the limit matches it to within $0.009$ bit.
\end{abstract}

\input{sections/intro}
\input{sections/setup}
\input{sections/theoremA}
\input{sections/theoremB}
\input{sections/typical}
\input{sections/weights}
\input{sections/experiments}
\input{sections/related}
\input{sections/discussion}

\section*{Impact Statement}
This paper presents work whose goal is to advance the field of Machine Learning. There are many potential societal consequences of our work, none which we feel must be specifically highlighted here.

\bibliography{refs}
\bibliographystyle{icml2026}

\newpage
\appendix
\onecolumn
\numberwithin{lemma}{section}
\numberwithin{table}{section}
\input{appendix/proofA}
\input{appendix/proofB}
\input{appendix/proofC}
\input{appendix/proofS}
\input{appendix/experiments}

\end{document}

%% file: authors.tex
\begin{icmlauthorlist}
  \icmlauthor{Runguo Li}{uiuc}
\end{icmlauthorlist}
\icmlaffiliation{uiuc}{University of Illinois Urbana-Champaign (UIUC)}
\icmlcorrespondingauthor{Runguo Li}{runguol2@illinois.edu}

%% file: sections/intro.tex
\section{Introduction}
\label{sec:intro}

Layerwise post-training quantization replaces a weight matrix $W$ by $\hat W$ so as to minimize the output error $\E\|(W-\hat W)x\|^2=\tr\big((W-\hat W)\Sigma(W-\hat W)^\top\big)$, where $\Sigma=\E[xx^\top]$ is estimated on calibration data. GPTQ~\citep{frantar2023gptq}, which quantizes one input column at a time and feeds the rounding error forward, has recently been shown to be exactly Babai's nearest-plane algorithm~\citep{babai1986lovasz} on the lattice generated by the Cholesky factor of $\Sigma$~\citep{chen2026geometry,birnick2026lattice}. Its quality is therefore governed by the Gram--Schmidt lengths of that basis, the Cholesky pivots $\pi_1,\dots,\pi_n$ of $\Sigma$.

Two remedies for the suboptimality of GPTQ are in use. \emph{Rotation} methods apply an orthogonal transform to weights and activations before running GPTQ, either random (QuIP, QuIP\#, QuaRot) or learned (SpinQuant)~\citep{chee2023quip,tseng2024quipsharp,ashkboos2024quarot,liu2025spinquant}. (An optimal data-dependent rotation can even reach the lattice bound~\citep{keung2026revisiting}; we study data-independent random rotations, as in the open problem and in practice.) \emph{Waterfilling} keeps the basis and assigns every column its own step size: WaterSIC~\citep{lifar2026watersic} reverse-waterfills the rate across columns. In the high-rate, entropy-coded regime the two can be compared exactly. WaterSIC is within $\tfrac12\log_2\frac{2\pi e}{12}\approx0.255$ bit of the rate-distortion limit for \emph{every} $\Sigma$, and GPTQ pays an additional
\begin{equation}
\Delta(\Sigma)=\tfrac12\log_2\frac{\AM(\pi_1,\dots,\pi_n)}{\GM(\pi_1,\dots,\pi_n)}
\label{eq:gap-intro}
\end{equation}
bits per weight~\citep{lifar2026watersic}. The gap $\Delta$ is unbounded over $\Sigma$, but on Llama-3-8B a random rotation brings it to about $0.1$ bit~\citep{ordentlich2026hrqmm2}. Whether rotation can replace waterfilling in general was left as an open problem:
\begin{quote}\small
``It is an interesting open problem to estimate worst possible gap (over possible spectra $\lambda_j\ge\epsilon$) between GPTQ with rotation and WaterSIC''~\citep[\S III-E]{ordentlich2026hrqmm2}.
\end{quote}
We answer this question in the setting in which it was posed (high rate, per-column entropy coding, Haar rotations), and explain why the gap is nonetheless small on real layers. Fixed-rate formats are outside this setting, and our conclusions should not be read as statements about INT4-style deployment (\cref{sec:exp}).

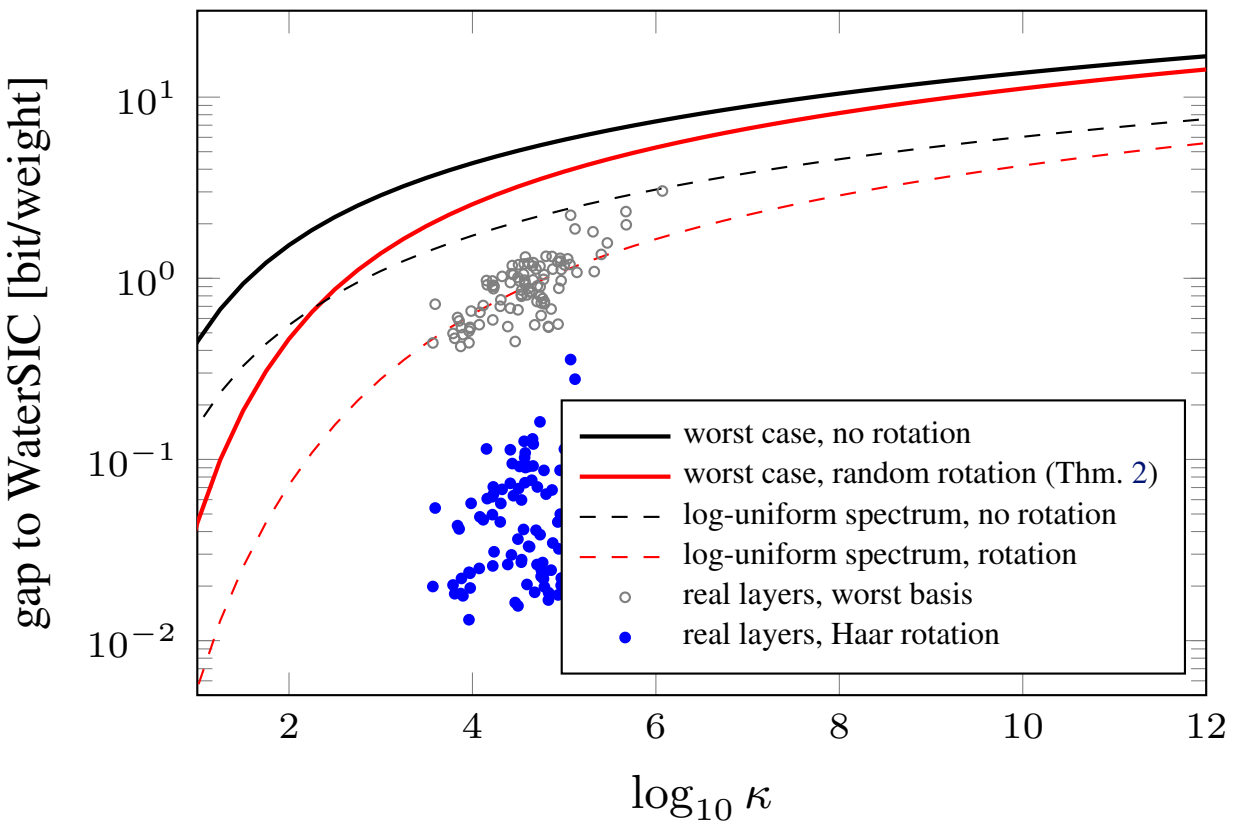
\begin{figure}[t]
\centering
\begin{tikzpicture}
\begin{semilogyaxis}[
  width=\columnwidth, height=0.74\columnwidth,
  xmin=1, xmax=12, ymin=0.005, ymax=30,
  xlabel={$\log_{10}\kappa$}, ylabel={gap to WaterSIC [bit/weight]},
  label style={font=\small}, tick label style={font=\scriptsize},
  legend style={font=\tiny, at={(0.98,0.03)}, anchor=south east, row sep=-2pt},
  legend cell align=left]
\addplot[black, thick] table[x=logkappa, y=specht]{figs/data/worst.dat};
\addlegendentry{worst case, no rotation}
\addplot[red, thick] table[x=logkappa, y=worst_rot]{figs/data/worst.dat};
\addlegendentry{worst case, random rotation (Thm.~\ref{thm:B})}
\addplot[black, dashed] table[x=logkappa, y=logu_none]{figs/data/worst.dat};
\addlegendentry{log-uniform spectrum, no rotation}
\addplot[red, dashed] table[x=logkappa, y=logu_rot]{figs/data/worst.dat};
\addlegendentry{log-uniform spectrum, rotation}
\addplot[only marks, mark=o, mark size=0.9pt, gray] table[x=logkappa, y=pca]{figs/data/e3_layers.dat};
\addlegendentry{real layers, worst basis}
\addplot[only marks, mark=*, mark size=0.9pt, blue] table[x=logkappa, y=haar]{figs/data/e3_layers.dat};
\addlegendentry{real layers, Haar rotation}
\end{semilogyaxis}
\end{tikzpicture}
\vskip -0.1in
\caption{High-rate rate gap of GPTQ to WaterSIC as a function of the condition number $\kappa$. Solid: worst case over spectra without rotation (the Specht ratio) and with a Haar rotation (\cref{thm:B}); the two differ only by $\tfrac12\log_2\ln\kappa+O(1)$. Dashed: the log-uniform spectrum (\cref{prop:logscale}). Dots: $96$ Hessians of Qwen3-1.7B, Qwen3-8B-Base and Llama-3.1-8B-Instruct (damping $10^{-2}$), in the worst (PCA) basis and after a Haar rotation (\cref{sec:exp}).}
\label{fig:worst}
\end{figure}

\paragraph{Contributions.}
\begin{enumerate}[leftmargin=*, itemsep=1pt, topsep=2pt]
\item \textbf{The limit of rotated GPTQ} (\cref{thm:A}). After a Haar rotation, the $k$-th GPTQ pivot converges almost surely to $h_\nu(k/n)$, uniformly in $k\ge\delta n$ and uniformly over spectra $\nu$ with bounded condition number. Here $h_\nu(t)=1/S_\nu(-t)$ is the harmonic mean of the free compression of $\nu$. Consequently $\Delta\to\frac12\log_2(D(\nu)/\GM(\nu))$ with $D(\nu)=\int_0^1h_\nu$. The formula for $h_\nu$ is classical in the theory of Lyapunov exponents~\citep{newman1986distribution,kargin2008lyapunov}; our contribution is its identification with GPTQ, a finite-$n$, per-pivot version that is uniform in the spectrum, and the functional $D$ that governs the rate.
\item \textbf{The worst case} (\cref{thm:B}), which solves the open problem. For condition number $\kappa$, the worst rotated gap is $\frac12\log_2\kappa-\log_2\ln\kappa+\frac12-\frac1{\ln2}+o(1)$ bits, and every extremal spectrum is a two-point spectrum with mass $\approx2/\ln\kappa$ on the top eigenvalue. Without rotation the worst gap is $\frac12\log_2\kappa-\frac12\log_2\ln\kappa-\frac1{2\ln 2}+o(1)$. In the worst case rotation therefore saves only $\frac12\log_2\ln\kappa+O(1)$ bits, and cannot replace waterfilling. The same holds for the normalization $\lambda_j\ge\epsilon\,\AM(\lambda)$ (\cref{thm:Bprime}). The proof rests on a log-convexity property of the first variation of $D$ that pushes extremal spectra to the endpoints (\cref{lem:logconvex}).
\item \textbf{Why real layers are benign} (\cref{sec:typical}). An $\eta$-fraction of outlier eigenvalues of size $\mu$ costs about $\eta\mu$ in the worst basis, but after rotation only $\Theta(\eta^2\mu)$, with constant exactly $\frac12$ when the outliers matter ($\eta^2\mu\to\infty$). In finite dimension, $r$ spikes are removed as long as $(r+\ln n)\sum_j\mu_j=o(n^2)$, a condition that is sharp up to the factor $r+\ln n$. By contrast, spectra spread on a logarithmic scale keep the leading term, and rotation again saves only $\frac12\log_2\ln\kappa+O(1)$. Together with \cref{thm:B} this describes when rotation suffices (\cref{sec:discussion}).
\item \textbf{Beyond Gaussian weights} (\cref{sec:weights}). The WaterSIC analysis models weights as i.i.d.\ Gaussian, under which rotations leave the weight entropy unchanged. For real weights, under per-column entropy coding, the gap of rotated GPTQ to WaterSIC is the Hessian term of \cref{thm:A} plus a weight-side cost $\Cw(Q)$, and both can be computed before quantization. $\Cw(Q)=\WS(W)-\WS(WQ)\le\WS(W)$, where $\WS$ is the classical transform-coding gain plus a non-Gaussianity term.
\item \textbf{Experiments} (\cref{sec:exp}) on $96$ real Hessians. The rotated gap of real layers has median $0.04$ bit and maximum $0.36$ bit, against $2$--$5$ bits in the worst case for their $\kappa$ (\cref{fig:worst}), and the limit of \cref{thm:A} matches it to within $0.009$ bit. In actual entropy-coded quantization, the high-rate rate prediction holds to $\le0.015$ bit at $4$ bits per weight on all but one of $64$ layers, and $\Cw$ is predicted from the weights alone with median error $0.0008$ bit.
\end{enumerate}

%% file: sections/setup.tex
\section{Setup}
\label{sec:setup}

\paragraph{Layerwise quantization and GPTQ.}
Let $W\in\R^{a\times n}$ be a weight matrix acting on inputs $x\in\R^n$ with second moment $\Sigma=\E[xx^\top]\in\mathbb S^n_{++}$, and let the distortion be $\dist=\frac{1}{na}\tr\big((W-\hat W)\Sigma(W-\hat W)^\top\big)$. Write $\Sigma=U^\top U$ with $U$ upper triangular. The \emph{pivots}
\begin{equation}
\pi_k:=U_{kk}^2=\frac{\det\Sigma^{(k)}}{\det\Sigma^{(k-1)}}
\label{eq:pivots}
\end{equation}
are the squared Gram--Schmidt lengths of the lattice basis, where $\Sigma^{(k)}$ is the leading $k\times k$ principal submatrix. Each pivot is the minimum residual second moment for linear prediction of $x_k$ from the preceding coordinates; for centered jointly Gaussian inputs it is also the conditional variance. GPTQ with a common step $\alpha$ is successive interference cancellation (Babai's nearest-plane algorithm) on this basis, with the coordinates ordered so that the first quantized column is the last Gram--Schmidt coordinate~\citep{chen2026geometry,birnick2026lattice,lifar2026watersic}. Note that $\prod_k\pi_k=\det\Sigma$ for every basis.

\paragraph{The rate gap.}
In the high-rate limit with per-column entropy coding and $a\to\infty$ output channels, \citet[Thm.~3.3]{lifar2026watersic} show that, at equal distortion, Huffman-GPTQ needs
\begin{equation}
\Delta(\Sigma):=\tfrac12\log_2\frac{\frac1n\sum_k\pi_k}{(\det\Sigma)^{1/n}}\ \ge0
\label{eq:gap}
\end{equation}
more bits per weight than WaterSIC. WaterSIC uses per-column steps $\alpha_k\propto U_{kk}^{-1}$ and is itself within $\frac12\log_2\frac{2\pi e}{12}$ bit of the waterfilling lower bound for every $\Sigma$. The denominator $\GM(\lambda)=(\det\Sigma)^{1/n}$ depends only on the eigenvalues $\lambda_1,\dots,\lambda_n$, while the numerator depends on the basis. Since $\sum_k\pi_k=\sum_j\lambda_j-\sum_{k<j}U_{kj}^2$~\citep{ordentlich2026hrqmm2}, the worst basis is the eigenbasis (PCA basis), where $\Delta=\frac12\log_2(\AM(\lambda)/\GM(\lambda))$; see also \citet[Thm.~4]{chee2023quip}. The same arithmetic-to-geometric mean ratio of successive-cancellation pivots appears in the analysis of zero-forcing decision feedback in communications~\citep{cioffi1995mmse,forney2003role}.

\paragraph{Rotations.}
Replacing $W\mapsto WV$ and $x\mapsto V^\top x$ for an orthogonal $V$ leaves the layer unchanged and replaces $\Sigma$ by $V^\top\Sigma V$. We study $V\sim\Haar(O(n))$; in practice randomized Hadamard transforms are used (\cref{sec:exp,sec:discussion}). The open problem asks for $\sup\Delta(V^\top\Sigma V)$ over spectra.

\paragraph{Normalization.}
Since $\Delta$ is invariant under scaling of $\Sigma$, the constraint ``$\lambda_j\ge\epsilon$'' needs a reference scale. Our main normalization is the condition number, \textbf{(N1)} $\supp\nu\subset[\epsilon,1]$, i.e.\ $\kappa(\Sigma)\le\kappa:=1/\epsilon$. We also treat the relative form \textbf{(N2)} $\lambda_j\ge\epsilon\cdot\AM(\lambda)$. Throughout, $\nu$ denotes a spectral measure, $\nu_n=\frac1n\sum_j\delta_{\lambda_j}$, $L:=\ln\kappa$, and $\AM,\GM,\HM$ are the arithmetic, geometric and harmonic means of $\nu$.

\paragraph{Basis gap and lattice gap.}
Column orderings and lattice reduction (LLL) change only the basis of the lattice; rotations and per-column steps change the lattice itself. On real LLM Hessians, full LLL reduction after a greedy ordering lowers $\sum_k\pi_k$ by at most $0.37\%$ (\cref{app:lll}). The gap~\eqref{eq:gap} is therefore essentially a property of the lattice, which is why we compare rotations with waterfilling.

%% file: sections/theoremA.tex
\section{The Limit of Rotated GPTQ}
\label{sec:A}

\begin{definition}[Pivot profile]
\label{def:h}
For a probability measure $\nu$ on $[\epsilon,1]$ and $s>0$, let
\[
T_\nu(s)=\int\frac{x}{s+x}\,d\nu(x),\qquad A_\nu(s)=\int\frac{d\nu(x)}{s+x}.
\]
$T_\nu$ decreases strictly from $1$ to $0$ on $(0,\infty)$. For $t\in(0,1)$ let $s_t$ be the root of $T_\nu(s)=t$ and set
\[
h_\nu(t):=\frac{t\,s_t}{1-t}=\frac{T_\nu(s_t)}{A_\nu(s_t)},
\]
with $h_\nu(0):=\AM(\nu)$ and $h_\nu(1):=\HM(\nu)$, and

\[
D(\nu):=\int_0^1h_\nu(t)\,dt,\qquad \Delta(\nu):=\tfrac12\log_2\frac{D(\nu)}{\GM(\nu)}.
\]
\end{definition}

\begin{theorem}[Rotated GPTQ pivots]
\label{thm:A}
Fix $\epsilon,\delta\in(0,1)$. For each $n$ let $\Lambda_n$ be diagonal with entries in $[\epsilon,1]$ and empirical spectral measure $\nu_n$, let $V_n\sim\Haar(O(n))$, and let $\pi_k$ be the pivots~\eqref{eq:pivots} of $V_n^\top\Lambda_nV_n$. Then, almost surely,
\begin{enumerate}[label=(\roman*), leftmargin=*, itemsep=0pt, topsep=1pt]
\item $\max_{\delta n\le k\le n}\big|\pi_k-h_{\nu_n}(k/n)\big|\to0$;
\item $\big|\frac1n\sum_{k=1}^n\pi_k-D(\nu_n)\big|\to0$;
\item $\big|\Delta(V_n^\top\Lambda_nV_n)-\Delta(\nu_n)\big|\to0$.
\end{enumerate}
The spectra need not converge. Moreover, for every $r>0$ there are constants $C,c>0$ depending only on $(\epsilon,\delta,r)$ such that the probability that any of the three quantities exceeds $r$ is at most $Cne^{-cn}$, uniformly over all spectra in $[\epsilon,1]$. If $\nu_n\Rightarrow\nu$, then $\nu_n$ may be replaced by $\nu$.
\end{theorem}

Three remarks on the statement. First, the first $\delta n$ pivots need no control: they are at most $1$ and contribute at most $\delta$ to the average. Second, uniformity over spectra is what lets us transfer the worst-case analysis of \cref{sec:B} to finite $n$ (\cref{cor:finite-n}). Third, the constants degrade as $\epsilon\to0$, so the bound is not quantitative for real layers with $\kappa\approx10^4$--$10^6$. The prediction is nevertheless accurate there (\cref{sec:exp}), consistent with real spectra being a narrow bulk plus a small fraction of large eigenvalues. That is the regime of \cref{prop:outliers,prop:spikes}. Computing $\Delta(\nu)$ takes one eigendecomposition of $\Sigma$, followed by a monotone scalar root-finding for $s_t$ at each point of a grid in $t$.

\begin{proposition}[Properties of the profile]
\label{prop:props}
For every $\nu$ on $[\epsilon,1]$:
\begin{enumerate}[label=(\alph*), leftmargin=*, itemsep=0pt, topsep=1pt]
\item $h_\nu(t)=1/S_\nu(-t)$, where $S_\nu$ is Voiculescu's $S$-transform. Equivalently, $h_\nu(t)$ is the harmonic mean of the free compression of $\nu$ by a projection of trace $t$.
\item $h_\nu$ is continuous and non-increasing from $\AM(\nu)$ to $\HM(\nu)$, and strictly decreasing unless $\nu$ is a point mass.
\item $\int_0^1\ln h_\nu(t)\,dt=\int\ln x\,d\nu(x)$, the limit of $\sum_k\ln\pi_k=\ln\det\Sigma$.
\item $\GM(\nu)\le D(\nu)\le\AM(\nu)$, with equality in either inequality only for point masses. Thus a Haar rotation is strictly better than the worst basis, but strictly worse than WaterSIC.
\item If $\nu\le_{\rm st}\nu'$ (stochastic order), then $h_\nu\le h_{\nu'}$ pointwise.
\end{enumerate}
\end{proposition}

\paragraph{Relation to prior work.}
The identity $h_\nu(t)=1/S_\nu(-t)$ says that $\frac12\ln\pi_k$ converges to the $k$-th Lyapunov exponent of a product of rotation-invariant matrices. In this form the limit is classical. \citet[Eqs.~(4) and~(11)]{newman1986distribution} gives $\E\ln\pi_k$ and its large-$N$ limit, and \citet[Thm.~2, Cors.~4--5]{kargin2008lyapunov} gives the operator-algebraic version, of which \cref{prop:props}(c) is Corollary~5. The same quantity appears as the per-stream SINR of MMSE successive interference cancellation with Haar spreading~\citep[Example~2.51, \S3.2.2.2]{tulino2004random}. Our additions are fourfold:
\begin{itemize}[leftmargin=*, itemsep=0pt, topsep=1pt]
\item the identification $\pi_k=R_{kk}^2$ for the QR factorization of $\Sigma^{1/2}V$ (so $R=U$ for $V^\top\Sigma V$), which links GPTQ to this theory;
\item convergence of the pivots themselves (not their logarithms), almost surely, uniformly in $k\ge\delta n$ and in the spectrum;
\item the arithmetic-mean functional $D(\nu)=\int_0^1dt/S_\nu(-t)$, which is what enters the rate;
\item an elementary proof that avoids free probability.
\end{itemize}
\citet[Eq.~(33)]{ordentlich2026hrqmm2} approximate the pivots by ratios of expected minors, $\pi_k\approx\frac{k}{n-k+1}\frac{e_k(\lambda)}{e_{k-1}(\lambda)}$ (an annealed approximation). We show that for $\delta n\le k\le(1-\delta)n$ its logarithm is within $O(\sqrt{\ln n/n})$ of $\ln h_{\nu_n}(k/n)$, deterministically and uniformly over spectra in $[\epsilon,1]$ (\cref{prop:annealed}). So on this range it is asymptotically exact, and the quenched and annealed limits coincide. We are not aware of a finite-$n$, per-pivot result that is uniform in the spectrum, nor of a worst-case analysis of $D(\nu)$; in fixed dimension, \citet{reddy2019equality} identifies Lyapunov exponents with $\E\log R_{kk}$.

\paragraph{Proof idea (\cref{app:A}).}
By the Schur complement, $\pi_k=1/[(\Sigma^{(k)})^{-1}]_{kk}$. Haar invariance turns this diagonal entry into $\sum_iw_i^2/\gamma_i$, with $w$ uniform on the sphere and independent of the eigenvalues $\gamma_i$ of $\Sigma^{(k)}$; Hanson--Wright~\citep{rudelson2013hanson} concentrates it around $\frac1k\tr(\Sigma^{(k)})^{-1}$. A Gaussian representation of the first $k$ Haar columns, two Sherman--Morrison identities and leave-one-out concentration show that this normalized trace is close to $1/h_{\nu_n}(k/n)$. Interlacing gives Lipschitz control in $k$, which upgrades pointwise convergence to uniform convergence.

\paragraph{The two-point spectrum.}
For $\nu_q=(1-q)\delta_\epsilon+q\delta_1$, equation $T_\nu(s)=t$ is quadratic, and
\[
\frac{(q-t(1+\epsilon))_+}{1-t}\le h_{\nu_q}(t)\le\frac{(q+\epsilon-t)_+}{1-t}+\sqrt{\frac{\epsilon t}{1-t}}.
\]
The first $\approx qn$ Gram--Schmidt coordinates carry conditional variances $\approx\frac{q-t}{1-t}\gg\epsilon$. A single step size cannot adapt to them, and this is the source of the worst case below.

%% file: sections/theoremB.tex
\section{The Worst Case: Rotation Cannot Replace Waterfilling}
\label{sec:B}

Let $\cP_\epsilon$ be the probability measures on $[\epsilon,1]$, $\Delta^*(\epsilon):=\sup_{\nu\in\cP_\epsilon}\Delta(\nu)$ the worst rotated gap, and $\Delta^*_\none(\epsilon):=\sup_{\nu}\frac12\log_2\frac{\AM(\nu)}{\GM(\nu)}$ the worst gap in the worst basis. Recall $L=\ln(1/\epsilon)=\ln\kappa$.

\begin{theorem}[Worst-case gap under (N1)]
\label{thm:B}
Let $0<\epsilon<1$.
\begin{enumerate}[label=(\roman*), leftmargin=*, itemsep=0pt, topsep=1pt]
\item \emph{(Extremal structure.)} $\Delta^*(\epsilon)$ is attained, and every maximizer is a two-point measure $\nu_q=(1-q)\delta_\epsilon+q\delta_1$ with $q\in(0,1)$. Hence
\[
\Delta^*(\epsilon)=\max_{q\in(0,1)}\frac{\ln D(\nu_q)+(1-q)L}{2\ln2}
\]
(numerically the maximizing $q$ is unique; we do not prove this).
\item \emph{(Asymptotics.)} As $\epsilon\to0$,
\[
\Delta^*(\epsilon)=\tfrac12\log_2\kappa-\log_2\ln\kappa+\tfrac12-\tfrac1{\ln2}+o(1).
\]
\item \emph{(No rotation.)} $\Delta^*_\none(\epsilon)=\tfrac12\log_2\kappa-\tfrac12\log_2\ln\kappa-\tfrac1{2\ln2}+o(1)$, the asymptotics of the Specht ratio~\citep{specht1960theorie}.
\end{enumerate}
Consequently, in the worst case random rotation saves
\[
\Delta^*_\none-\Delta^*=\tfrac12\log_2\ln\kappa-\tfrac12+\tfrac1{2\ln2}+o(1)
\]
bits, while the leading term $\frac12\log_2\kappa$ is unchanged.
\end{theorem}

\begin{corollary}[Finite $n$]
\label{cor:finite-n}
If $\kappa(\Sigma_n)\le\kappa$ and $V_n\sim\Haar(O(n))$, then $\limsup_n\Delta(V_n^\top\Sigma_nV_n)\le\Delta^*(\epsilon)$ almost surely. Moreover, $\sup_{\kappa(\Sigma_n)\le\kappa}\Prob\big(\Delta(V_n^\top\Sigma_nV_n)>\Delta^*(\epsilon)+r\big)\to0$ for every $r>0$. Equality is attained by spectra at the quantiles of $\nu_{q^*}$.
\end{corollary}

\begin{table}[t]
\caption{Worst-case gaps in bits (exact, by one-dimensional maximization over $q$; see \cref{app:B-numerics}). ``Saved'' is $\Delta^*_\none-\Delta^*$.}
\label{tab:worst}
\vskip 0.05in
\centering
\small
\setlength{\tabcolsep}{4pt}
\begin{tabular}{lccccc}
\toprule
$\kappa$ & $\Delta^*_\none$ & $\Delta^*$ & asympt.\ (ii) & $q^*$ & saved\\
\midrule
$10^{2}$ & 1.53 & 0.46 & 0.18 & 0.39 & 1.06\\
$10^{4}$ & 4.32 & 2.56 & 2.50 & 0.225 & 1.76\\
$10^{6}$ & 7.35 & 5.27 & 5.24 & 0.149 & 2.08\\
$10^{12}$ & 16.82 & 14.22 & 14.20 & 0.073 & 2.60\\
$10^{48}$ & 75.61 & 72.00 & 72.00 & 0.018 & 3.61\\
\bottomrule
\end{tabular}
\vskip -0.1in
\end{table}

\paragraph{Proof idea (\cref{app:B}).}
Write $J(\nu)=\ln D(\nu)-\int\ln x\,d\nu$, so that $\Delta=J/(2\ln 2)$. $J$ is continuous on the compact set $\cP_\epsilon$, so a maximizer exists, and it is not a point mass since $J=0$ there. The Gateaux derivative of $D$ in the direction $\mu-\nu$ is $\int\Phi_\nu\,d(\mu-\nu)$, where
\[
\Phi_\nu(y)=-\int_0^1\frac{\bar h_\nu(s)}{s+y}ds+\int_1^\infty\frac{\bar h_\nu(s)\,y}{s(s+y)}ds
\]
and $\bar h_\nu=T_\nu/A_\nu$.
The first-order (KKT) condition says that every maximizer $\nu^*$ is supported on the maximizers over $[\epsilon,1]$ of $\psi(y)=\Phi_{\nu^*}(y)/D(\nu^*)-\ln y$. The crux is the following lemma.

\begin{lemma}[Log-convexity]
\label{lem:logconvex}
If $\nu$ is not a point mass, then $u\mapsto\Phi_\nu(e^u)$ is strictly convex. More precisely,
\[
\frac{d}{du}\Phi_\nu(e^u)=\int_\R\bar h_\nu(e^v)\,\ell(u-v)\,dv,
\]
with $\ell(r)=e^r/(1+e^r)^2$. That is, the derivative is the convolution of the strictly increasing function $v\mapsto\bar h_\nu(e^v)$ with the logistic density.
\end{lemma}

Hence $\psi\circ\exp$ is strictly convex on $[\ln\epsilon,0]$, its maximum over the interval is attained only at the endpoints, and every maximizer lives on $\{\epsilon,1\}$. In the natural coordinate $y$ the function $\Phi_\nu$ is \emph{concave}, which hides this structure: the right coordinate is $\ln y$. (Without rotation the first variation is $y/\AM-\ln y$, which is convex in both coordinates; this is the KKT proof of Specht's theorem.) The asymptotics (ii) follow from explicit bounds on $D(\nu_q)$, namely $\frac{q^2}{2}\lesssim D(\nu_q)\lesssim\frac{q^2}{2(1-q)}$, with the optimum at $q^*\approx2/L$.

\paragraph{The alternative normalization.}
The open problem does not specify the scale. Under (N2) the large eigenvalues are no longer capped at $\lambda_{\max}$, and the extremal structure changes, but the conclusion does not.

\begin{theorem}[Worst case under (N2)]
\label{thm:Bprime}
Under $\lambda_j\ge\epsilon\,\AM(\lambda)$:
\begin{enumerate}[label=(\roman*), leftmargin=*, itemsep=0pt, topsep=1pt]
\item for every truncation $\lambda_{\max}\le R\,\lambda_{\min}$, every maximizer of $\Delta$ is a two-point measure, and $\Delta^*_{N2}(\epsilon)$ is the supremum of $\Delta$ over two-point measures that satisfy (N2);
\item $\Delta^*_{N2}(\epsilon)=\tfrac12\log_2\tfrac1\epsilon-\tfrac12\log_2\ln\tfrac1\epsilon-\tfrac12-\tfrac1{2\ln2}+o(1)$.
\end{enumerate}
Without rotation the supremum is $\frac12\log_2\frac1\epsilon$, so the saving is $\frac12\log_2\ln\frac1\epsilon+\frac12+\frac1{2\ln2}+o(1)$. Numerically, the optimal two-point spectra have mass $q^*\approx1/L$ on an upper atom near $\AM/q^*$, and the mean constraint is tight, so that $\lambda_{\min}=\epsilon\,\AM$.
\end{theorem}

The proof (\cref{app:Bprime}) adds a Lagrange multiplier $-\beta y$ for the mean constraint. The KKT function is then no longer convex in $\ln y$, but a unimodality property of the smoothed derivative (\cref{lem:unimodal}) still limits it to at most two maximizers, one of them at the left endpoint.

\paragraph{Interpretation.}
Under both normalizations the worst rotated gap grows like $\frac12\log_2\kappa$, as without rotation. In the extremal spectra a fraction $\approx2/\ln\kappa$ (under (N2), $\approx1/\ln\kappa$) of the directions carries almost all the energy. After rotation these directions occupy the first $\approx qn$ GPTQ columns, and a common step size cannot allocate rate to them; WaterSIC's per-column steps can. Real layers sit far below it (\cref{fig:worst,sec:typical}).

%% file: sections/typical.tex
\section{Typical Spectra: Outliers versus Spread}
\label{sec:typical}

Real Hessian spectra have a narrow bulk plus a few very large eigenvalues (\cref{sec:exp}). Here we quantify what rotation does to each part. Proofs are in \cref{app:C}.

\paragraph{Outliers cost $\eta^2\mu$ after rotation.}
Let $\rho,\alpha$ be probability measures on $[\epsilon,b]$ and $\beta$ a compactly supported measure on $[\epsilon,\infty)$, and compare
\[
\nu=(1-\eta)\rho+\eta\alpha,\qquad \nu'=(1-\eta)\rho+\eta\beta .
\]
That is, $\nu'$ moves a fraction $\eta$ of the spectrum to arbitrary positions.

\begin{proposition}[Outliers]
\label{prop:outliers}
For $\eta\le\frac12$:
\begin{enumerate}[label=(\roman*), leftmargin=*, itemsep=0pt, topsep=1pt]
\item $\big|1/h_\nu(t)-1/h_{\nu'}(t)\big|\le\eta/(\epsilon t)$ for all $t\in(0,1]$;
\item $h_{\nu'}(t)\le bt/(t-\eta)$ for $t>\eta$;
\item $|D(\nu')-D(\nu)|\le\frac{4b^2\eta}{\epsilon}\ln\frac{1}{2\eta}+2\eta\big(\AM(\nu)+\AM(\nu')\big)$, where $2\eta\AM(\nu')\le2\eta b+2\eta^2\AM(\beta)$;
\item if $\nu=\rho$ and $\beta=\delta_\mu$, then $D(\nu')-D(\nu)=\frac12\eta^2\mu\,(1+o(1))$ as $\eta\to0$ and $\eta^2\mu\to\infty$, with $\rho$ fixed.
\end{enumerate}
\end{proposition}

In the worst basis the same outliers raise the gap through $\AM(\nu')-\AM(\nu)=\eta(\AM(\beta)-\AM(\alpha))$, which is first order in $\eta$. A rotation turns this cost of $\approx\eta\mu$ into one of order $\eta^2\mu$, with constant exactly $\frac12$ in the regime where the outliers matter. The reason is that the first $\approx\eta n$ GPTQ columns successively absorb the outlier subspace: column $k$ still sees outlier energy $\approx\mu(\eta n-k)/n$, and summing gives $\mu\eta^2n/2$. Rotation is therefore effective roughly when $\eta^2\mu\ll D(\nu)$, that is, when $\eta\ll\mu^{-1/2}$ on the scale of the bulk. The extremal spectra of \cref{thm:B}, with $\eta\approx2/\ln\kappa$, are far outside this regime. With bulk on $[\frac12,1]$ and $\mu=5\cdot10^5$, the leading-order estimate $\frac12\log_2\big(1+\frac{\eta^2\mu/2}{D}\big)$ gives $0.0024$ and $0.21$ bit for $\eta=10^{-4},10^{-3}$, against exact rotated gaps $0.003$ and $0.21$ (without rotation: $3.05$ and $4.70$).

\paragraph{Finitely many spikes of any size.}
\Cref{prop:outliers} is a limit statement. Massive activations, however, are a few directions whose size may grow with the dimension.
\begin{proposition}[Spikes]
\label{prop:spikes}
Suppose $\Sigma_n$ has $n-r_n$ eigenvalues in $[\epsilon,1]$ and $r_n$ eigenvalues $\mu_j\ge1$, and let $\nu^0_n$ be its spectrum with the spikes replaced by $1$. After a Haar rotation:
\begin{itemize}[leftmargin=*, itemsep=0pt, topsep=1pt]
\item if $r_n=o(n)$, the pivots with $k\ge\delta n$ still follow $h_{\nu^0_n}(k/n)$, whatever the sizes $\mu_j$;
\item if $(r_n+\ln n)\sum_j\mu_j=o(n^2)$, then $\Delta=\Delta(\nu^0_n)-\frac{1}{2n\ln2}\sum_j\ln\mu_j+o(1)$;
\item if $r_n=o(n)$, then $\sum_j\mu_j=o(n^2)$ is necessary, because the first pivot alone is of order $\sum_j\mu_j/n$.
\end{itemize}
\end{proposition}
After rotation the spikes can therefore only lower the gap, through the geometric mean. On our $96$ layers, with spikes defined as eigenvalues above $100\times$ the median and normalized by the largest remaining one, $(r_n+\ln n)\sum_j\mu_j/n^2\le0.03$ (median $9\cdot10^{-5}$; $\le0.2$ at a $30\times$ threshold). The proof's error term is a constant multiple of this quantity, so it indicates the regime; \cref{fig:e3} is the direct check.

\paragraph{Spread on a logarithmic scale is not removed.}
\begin{proposition}[Log-scale spectra]
\label{prop:logscale}
Let $\nu_\kappa$ be the law of $\kappa^{-U}$, where $U$ has a bounded density $f$ on $[0,1]$ that is continuous at $0$ with $f(0)>0$. As $\kappa\to\infty$, $(\ln\kappa)^2D(\nu_\kappa)\to f(0)^2\zeta(2)$, and
\begin{align*}
\Delta_{\rot}&=\E[U]\cdot\tfrac12\log_2\kappa-\log_2\ln\kappa\\
&\qquad+\log_2f(0)+\tfrac12\log_2\zeta(2)+o(1),\\
\Delta_{\none}&=\E[U]\cdot\tfrac12\log_2\kappa-\tfrac12\log_2\ln\kappa\\
&\qquad+\tfrac12\log_2f(0)+o(1).
\end{align*}
\end{proposition}

The leading term is again unchanged, and rotation saves $\frac12\log_2\ln\kappa+O(1)$, the same order as in the worst case. For the log-uniform spectrum ($f\equiv1$) the profile is explicit:
\[
h(t)=\frac{t\,(1-e^{-(1-t)L})}{(1-t)(e^{tL}-1)}.
\]
This gives $\Delta_\rot=\frac14\log_2\kappa-\log_2L+0.359+\frac{1.054}{L}+\frac{2.077}{L^2}+O(L^{-3})$ (\cref{app:C}); the dashed curves in \cref{fig:worst} are the exact values. Spectra whose mass sits near the top, such as $x=\epsilon+(1-\epsilon)u^\beta$, saturate instead.

%% file: sections/weights.tex
\section{Beyond Gaussian Weights: the Weight-Side Cost of Rotation}
\label{sec:weights}

The high-rate analysis of WaterSIC models the rows of $W$ as i.i.d.\ $\mathcal N(0,\sigma_W^2I)$~\citep[\S3.1]{lifar2026thesis}. Under this model $W$ and $WQ$ have the same distribution, and WaterSIC is invariant to rotations. Real weights are not isotropic Gaussian, and with per-column entropy coding a rotation also changes the code length of the weights. The variance part of this effect is the classical coding gain of transform coding~\citep{huang1963block,goyal2001theoretical}. What is specific to our setting is how it combines with the Hessian term of \cref{thm:A}.

Let the rows of $W$ be i.i.d.\ with a density on $\R^n$, and let $\sigma_i^2$ be the variance of column $W_i$. Write $\hd(\cdot)$ for differential entropy in bits and $\hd_{\rm col}(W)=\frac1n\sum_i\hd(W_i)$.

\begin{assumption}[High-rate column entropy]
\label{ass:H}
For a step matrix $\alpha\,\diag(\tilde\alpha_i)$ with $\prod_i\tilde\alpha_i=1$, as $\alpha\to0$ the distortion of successive interference cancellation satisfies $\dist/\alpha^2\to\frac1{12n}\sum_i\tilde\alpha_i^2\pi_i$, and the entropy of the $i$-th integer column satisfies $H(Z_i)-\log_2\frac1\alpha\to \hd(W_i)-\log_2\tilde\alpha_i$.
\end{assumption}

\citet[Thm.~3.3]{lifar2026watersic} prove \cref{ass:H} for i.i.d.\ Gaussian rows. For general rows, the rounding offsets of SIC depend on the other columns, and we take the assumption as given; our experiments support it to $\approx0.001$ bit (\cref{sec:exp}).

\begin{proposition}[Weight-side cost]
\label{prop:S}
Let $Q$ be orthogonal and suppose \cref{ass:H} holds before and after rotation. At equal distortion, as $\dist\to0$,
\begin{align*}
R_{\rm PWSIC}(WQ)-R_{\rm PWSIC}(W)&\to \Cw(Q),\\
R_{\rm GPTQ}(WQ)-R_{\rm PWSIC}(W)&\to\Delta(Q^\top\Sigma Q)+\Cw(Q).
\end{align*}
Here $R(WQ)$ is the rate for the rotated pair $(WQ,Q^\top\Sigma Q)$, PWSIC denotes PlainWaterSIC, and $\Cw(Q):=\hd_{\rm col}(WQ)-\hd_{\rm col}(W)$. Moreover $\Cw(Q)=\WS(W)-\WS(WQ)$ exactly, where
\[
\WS(W):=\tfrac12\log_2\frac{\AM(\sigma_1^2,\dots,\sigma_n^2)}{\GM(\sigma_1^2,\dots,\sigma_n^2)}+\frac1n\sum_i\Neg(W_i)\ \ge0
\]
and $\Neg(X)=\frac12\log_2(2\pi e\,\mathrm{Var}X)-\hd(X)$ is the negentropy. In particular $\Cw(Q)\le\WS(W)$, with equality iff all columns of $WQ$ are Gaussian with equal variances.
\end{proposition}

The identity is elementary: it holds because $\sum_i\sigma_i^2=\tr\mathrm{Cov}$ is rotation invariant. The operational content of \cref{prop:S} is the decomposition into a Hessian term and a weight term, both computable before quantization. A rotation trades the column non-uniformity of $W$ for that of $WQ$:
\begin{itemize}[leftmargin=*, itemsep=0pt, topsep=1pt]
\item The first term of $\WS$ is the transform-coding gain of the native column basis, which a variance-equalizing rotation gives up.
\item The second term is non-Gaussianity, which random projections tend to remove~\citep{diaconis1984asymptotics}. The same Gaussianization by rotation is used for entropy-coding KV caches by \citet{su2026dquant}.
\end{itemize}
Two consequences follow:
\begin{itemize}[leftmargin=*, itemsep=0pt, topsep=1pt]
\item PlainWaterSIC's Hessian side is already basis free, so a rotation changes only its weight side, and equalizing the columns costs rate. We measured $\Cw>0$ on all $64$ layers.
\item For GPTQ, rotation helps under entropy coding exactly when the Hessian-side saving $\frac12\log_2\frac{\AM(\pi)}{\AM(\pi^Q)}$ exceeds $\Cw(Q)$. Both terms can be computed before quantization.
\end{itemize}

%% file: sections/experiments.tex
\section{Experiments}
\label{sec:exp}

\paragraph{Setup.}
We use Qwen3-1.7B, Qwen3-8B-Base~\citep{qwen3} and Llama-3.1-8B-Instruct~\citep{llama3}. From $8$ evenly spaced blocks per model we take the Hessians of the four distinct linear-layer inputs (shared q/k/v input, o\_proj input, shared gate/up input, down\_proj input). This gives $96$ matrices with $n=2048$--$14336$, computed in float64 from $128$ WikiText-2~\citep{merity2017pointer} training sequences of $2048$ tokens. Unless stated otherwise the damping is $10^{-2}\cdot\overline{\diag\Sigma}$, the GPTQ default. Rotations are Haar, or randomized Hadamard $DH_n$ (one round) and $D_1H_nD_2H_n$ (two rounds), each with $3$ seeds. Synthetic experiments, other dampings and all further details are in \cref{app:exp}.

\paragraph{\Cref{thm:A} on synthetic spectra.}
On five synthetic spectra, $|\Delta_n-\Delta(\nu)|\le0.017$ bit at $n=256$ and $\le0.0004$ bit at $n=8192$ under Haar rotations (\cref{app:exp}). Numerical maximization over measures with up to $8$ atoms never beat the best two-point spectrum.

\begin{figure}[t]
\centering
\begin{tikzpicture}
\begin{axis}[
  width=0.98\columnwidth, height=0.62\columnwidth,
  xmin=0, xmax=0.38, ymin=0, ymax=0.38,
  xlabel={limit $\Delta(\nu_{\rm emp})$ of \cref{thm:A} [bit]},
  ylabel={finite-$n$ gap [bit]},
  label style={font=\small}, tick label style={font=\scriptsize},
  legend style={font=\tiny, at={(0.03,0.97)}, anchor=north west, row sep=-2pt},
  legend cell align=left]
\addplot[only marks, mark=*, mark size=1pt, blue] table[x=pred, y=haar, restrict expr to domain={\thisrow{model}}{1:1}]{figs/data/e3_layers.dat};
\addlegendentry{Qwen3-1.7B, Haar}
\addplot[only marks, mark=*, mark size=1pt, orange] table[x=pred, y=haar, restrict expr to domain={\thisrow{model}}{2:2}]{figs/data/e3_layers.dat};
\addlegendentry{Qwen3-8B-Base, Haar}
\addplot[only marks, mark=*, mark size=1pt, teal] table[x=pred, y=haar, restrict expr to domain={\thisrow{model}}{3:3}]{figs/data/e3_layers.dat};
\addlegendentry{Llama-3.1-8B-Instruct, Haar}
\addplot[only marks, mark=o, mark size=1.4pt, gray] table[x=pred, y=had1]{figs/data/e3_layers.dat};
\addlegendentry{all, randomized Hadamard (1 round)}
\addplot[only marks, mark=x, mark size=1.4pt, red] table[x=pred, y=had2]{figs/data/e3_layers.dat};
\addlegendentry{all, randomized Hadamard (2 rounds)}
\addplot[black, dashed, domain=0:0.38] {x};
\end{axis}
\end{tikzpicture}
\vskip -0.1in
\caption{Finite-$n$ high-rate gap~\eqref{eq:gap} of rotated GPTQ, computed from the Cholesky pivots of each rotated Hessian, versus the limit of \cref{thm:A} computed from its empirical spectrum. $96$ real Hessians, damping $10^{-2}$, mean over $3$ seeds. Haar: every deviation is positive and at most $0.006$ bit at this damping ($0.009$ bit over all three). Hadamard (one or two rounds) deviates mainly on o\_proj inputs.}
\label{fig:e3}
\end{figure}
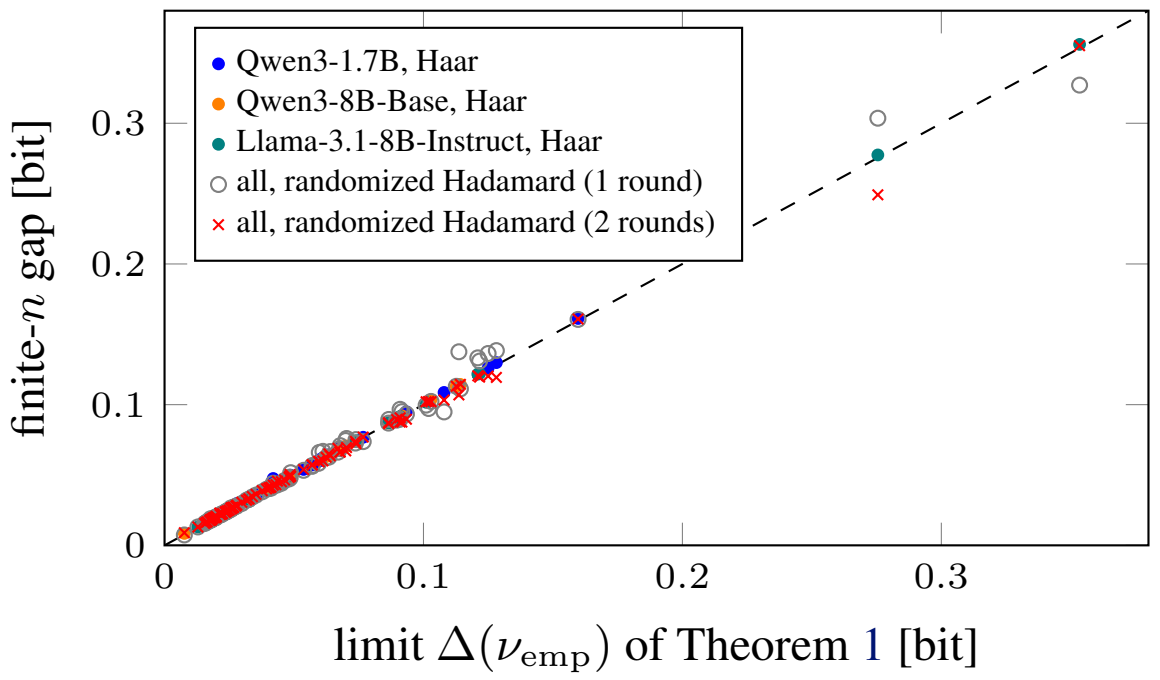

\paragraph{\Cref{thm:A} on real Hessians.}
Here the gap is computed exactly from the pivots of the rotated finite matrix. Real spectra violate the bounded-support assumption: the largest eigenvalue is $10^3$--$10^5$ times the median. The limit nevertheless holds (\cref{fig:e3}), as \cref{prop:spikes} explains. Over all $270$ (layer, damping) pairs, the Haar gap exceeds $\Delta(\nu_{\rm emp})$ by between $5\cdot10^{-5}$ and $0.009$ bit. At damping $10^{-2}$, randomized Hadamard is within $0.01$ bit of the prediction on $93\%$ of the layers with one round and $99\%$ with two. The main exception is the o\_proj input, whose covariance is blocked by attention head: one round deviates by $-0.013$ to $+0.028$ bit, and two rounds give gaps up to $0.026$ bit \emph{below} the limit, which a random input permutation removes (\cref{app:e3}). On synthetic axis-aligned spectra, one round gives gaps $0.02$--$0.15$ bit below the limit, while two rounds are indistinguishable from Haar.

\paragraph{Typical versus worst case.}
After a Haar rotation, the median gap is $0.041$ bit, $85\%$ of the layers are below $0.1$ bit, and the maximum is $0.356$ bit (Llama block~0, q/k/v input). The worst case $\Delta^*(\kappa)$ at each layer's own condition number is $2$--$5$ bits (\cref{fig:worst}). (The ratio, median $1.2\%$, depends on $\kappa$, hence on the damping.) Rotation saves a median of $0.15$ bit over unrotated GPTQ and $0.85$ bit over the worst (PCA) basis. Removing the largest $1\%$ of eigenvalues reduces the PCA-basis gap by $30$--$60\%$ but the rotated gap by only $0.005$--$0.02$ bit, consistent with the $\eta^2\mu$ scaling of \cref{prop:outliers}. One effect lies outside our theory: unrotated GPTQ with act-order beats a random rotation on $27\%$ of the layers, including $19$ of the $24$ gate/up inputs. Axis-aligned outliers do not explain this: there the top eigenvectors are no more concentrated than elsewhere, and all $26$ layers still win after clipping the largest $1\%$ of eigenvalues (\cref{app:e3}).

\begin{figure*}[t]
\centering
\input{figs/fig_e4}
\vskip -0.1in
\caption{Left and middle: measured equal-distortion rate gap of entropy-coded GPTQ to PlainWaterSIC versus the high-rate prediction, without rotation (left) and in the Hadamard basis (middle, same-basis gap), on $64$ layers at $R=2,3,5$; the dashed line is $y=x$. Right: WikiText-2 perplexity at equal entropy-coded rate (lines; blue Qwen3-1.7B, orange Llama-3.1-8B-Instruct; seed $0$) and of fixed-rate INT-$b$ g128 with and without Hadamard (crosses; rates include scales). Dotted horizontal lines: bf16.}
\label{fig:e4}
\end{figure*}
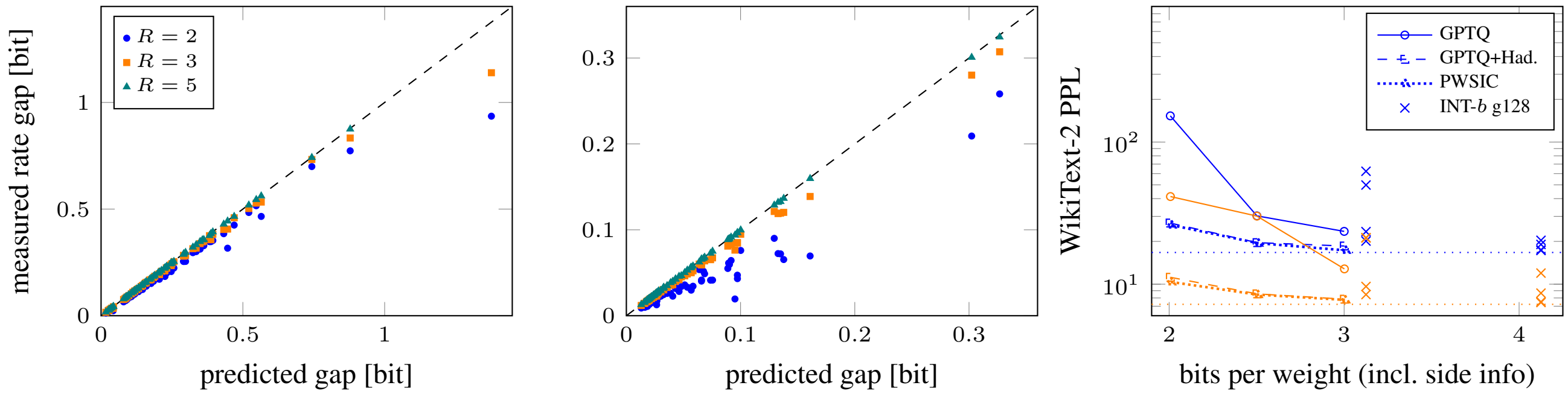

\paragraph{Rate side and weight-side cost.}
We quantized $64$ layers of Qwen3-1.7B and Llama-3.1-8B-Instruct with entropy-coded GPTQ and PlainWaterSIC. Rates are ideal per-column empirical entropies (\cref{app:e4}).
\begin{itemize}[leftmargin=*, itemsep=0pt, topsep=1pt]
\item \emph{Prediction accuracy.} The equal-distortion rate gap matches the high-rate prediction~\eqref{eq:gap} at $R=4$, in the native basis, with act-order and after Hadamard (\cref{tab:e4a}; \cref{fig:e4} shows $R=2,3,5$). The median error is $\le0.003$ bit and the maximum $0.015$ bit, except on Llama block~0's q/k/v input (near-dead input features), where it is $0.07$ bit. The per-layer rank correlation is $1.00$ for $R\ge3$. At $R=2$ the measured gaps are $0.02$--$0.03$ bit smaller than predicted.
\item \emph{Weight-side cost.} Rotating PlainWaterSIC costs $\Cw$, with median $0.005$--$0.036$ bit per layer type and $1.36$ bit on that Llama layer. The weights-only estimate of \cref{prop:S} predicts $\Cw$ with median error $0.0008$ bit and maximum error $0.034$ bit. On that layer Hadamard makes entropy-coded GPTQ \emph{worse} ($1.31\to1.68$ bit), because $\Cw$ exceeds the Hessian-side saving; on the other $63$ layers rotation helps.
\end{itemize}

\paragraph{End to end.}
We quantize every linear layer of all decoder blocks with sequential GPTQ and report WikiText-2 perplexity (\cref{tab:ppl}, \cref{fig:e4}). As the layerwise gaps predict, GPTQ without rotation is far worse at $2$--$3$ bits on both models. PlainWaterSIC without rotation and GPTQ with Hadamard are close. In the one setting with three calibration seeds (Qwen, $3$ bits), PlainWaterSIC was better for each seed and much more stable ($17.33\pm0.02$ versus $18.80\pm1.14$). This is an observation; more seeds and models would be needed to generalize it. In the single-seed settings the differences are within the seed-to-seed spread of GPTQ with Hadamard.

\begin{table}[t]
\caption{WikiText-2 perplexity (seqlen 2048) at equal entropy-coded rate. Entries with $\pm$: mean $\pm$ s.d.\ over 3 calibration seeds; other entries: seed 0 only. bf16: Qwen 16.72, Llama 7.22.}
\label{tab:ppl}
\vskip 0.05in
\centering
\small
\setlength{\tabcolsep}{3.5pt}
\begin{tabular}{llccc}
\toprule
model & bits & GPTQ & GPTQ+Had. & PWSIC\\
\midrule
Qwen3-1.7B & 2 & 152.95 & 26.93 & 26.06\\
 & 3 & 23.53 & 18.80$\pm$1.14 & 17.33$\pm$0.02\\
Llama-3.1-8B-I. & 2 & 41.37 & 11.18 & 10.37\\
 & 3 & 12.81 & 7.85 & 7.76\\
\bottomrule
\end{tabular}
\vskip -0.1in
\end{table}

\paragraph{Scope: fixed-rate formats.}
With fixed-rate INT-$b$ and group-128 scales the theory does not apply. There, WaterSIC's column multipliers give no gain ($\le0.008$ bit-equivalent), the format itself costs $0.8$--$1.0$ bit relative to entropy coding, and rotation helps mainly through clipping (\cref{app:e4}). After a Haar rotation, column ordering is asymptotically irrelevant: the saving from greedy min-pivot ordering decays like $n^{-1/2}$ (\cref{app:ordering}).

%% file: figs/fig_e4.tex
\begin{tikzpicture}
\begin{groupplot}[
  group style={group size=3 by 1, horizontal sep=1.25cm},
  width=0.355\textwidth, height=0.29\textwidth,
  label style={font=\small}, tick label style={font=\scriptsize}, title style={font=\small},
  legend style={font=\tiny, row sep=-2pt}, legend cell align=left]
\nextgroupplot[xmin=0, xmax=1.45, ymin=0, ymax=1.45,
  xlabel={predicted gap [bit]}, ylabel={measured rate gap [bit]},
  legend style={at={(0.03,0.97)}, anchor=north west}]
\addplot[only marks, mark=*, mark size=0.9pt, blue] table[x=pred_none, y=none2]{figs/data/e4_gaps.dat};
\addlegendentry{$R=2$}
\addplot[only marks, mark=square*, mark size=0.9pt, orange] table[x=pred_none, y=none3]{figs/data/e4_gaps.dat};
\addlegendentry{$R=3$}
\addplot[only marks, mark=triangle*, mark size=1.1pt, teal] table[x=pred_none, y=none5, unbounded coords=discard]{figs/data/e4_gaps.dat};
\addlegendentry{$R=5$}
\addplot[black, dashed, domain=0:1.45] {x};
\nextgroupplot[xmin=0, xmax=0.36, ymin=0, ymax=0.36,
  xlabel={predicted gap [bit]}, scaled ticks=false,
  xticklabel style={/pgf/number format/fixed}, yticklabel style={/pgf/number format/fixed}]
\addplot[only marks, mark=*, mark size=0.9pt, blue] table[x=pred_had1, y=had2]{figs/data/e4_gaps.dat};
\addplot[only marks, mark=square*, mark size=0.9pt, orange] table[x=pred_had1, y=had3]{figs/data/e4_gaps.dat};
\addplot[only marks, mark=triangle*, mark size=1.1pt, teal] table[x=pred_had1, y=had5]{figs/data/e4_gaps.dat};
\addplot[black, dashed, domain=0:0.36] {x};
\nextgroupplot[ymode=log, xmin=1.9, xmax=4.25, ymin=6, ymax=900,
  xlabel={bits per weight (incl.\ side info)}, ylabel={WikiText-2 PPL},
  legend style={at={(0.98,0.98)}, anchor=north east}]
\addplot[blue, mark=o, mark size=1.2pt] table[x=xh, y=yh]{figs/data/e4_ppl_qwen.dat};
\addlegendentry{GPTQ}
\addplot[blue, dashed, mark=square, mark size=1.2pt] table[x=xr, y=yr]{figs/data/e4_ppl_qwen.dat};
\addlegendentry{GPTQ+Had.}
\addplot[blue, densely dotted, thick, mark=triangle, mark size=1.4pt] table[x=xw, y=yw]{figs/data/e4_ppl_qwen.dat};
\addlegendentry{PWSIC}
\addplot[only marks, mark=x, mark size=2pt, blue] table[x=bits, y=ppl]{figs/data/e4_int_qwen.dat};
\addlegendentry{INT-$b$ g128}
\addplot[orange, mark=o, mark size=1.2pt] table[x=xh, y=yh]{figs/data/e4_ppl_llama.dat};
\addplot[orange, dashed, mark=square, mark size=1.2pt] table[x=xr, y=yr]{figs/data/e4_ppl_llama.dat};
\addplot[orange, densely dotted, thick, mark=triangle, mark size=1.4pt] table[x=xw, y=yw]{figs/data/e4_ppl_llama.dat};
\addplot[only marks, mark=x, mark size=2pt, orange] table[x=bits, y=ppl]{figs/data/e4_int_llama.dat};
\addplot[blue, thin, dotted, domain=1.9:4.25] {16.7167};
\addplot[orange, thin, dotted, domain=1.9:4.25] {7.2172};
\end{groupplot}
\end{tikzpicture}

%% file: sections/related.tex
\section{Related Work}
\label{sec:related}

\paragraph{GPTQ as lattice decoding.}
\citet{chen2026geometry} and \citet{birnick2026lattice} proved that GPTQ~\citep{frantar2023gptq} is Babai's nearest-plane algorithm~\citep{babai1986lovasz}. \citet{chen2026geometry} also give error bounds without clipping and the min-pivot ordering, and \citet{birnick2026lattice} proposes an LLL-based variant~\citep{lenstra1982factoring}. The same view underlies randomized and box-constrained decoders~\citep{wang2026ojbkq} and finite-sample analyses of OPTQ~\citep{zhang2025provable}.

\paragraph{Information-theoretic PTQ.}
\citet{ordentlich2024optimal,ordentlich2026hrqmm1,ordentlich2026hrqmm2} develop the rate-distortion theory of quantized matrix products. WaterSIC~\citep{lifar2026watersic} attains the waterfilling bound up to $0.255$ bit for every input covariance, and shows that the GPTQ gap~\eqref{eq:gap} can be arbitrarily large. We answer the open problem of \citet{ordentlich2026hrqmm2}. A recent survey~\citep{jokar2026transforms} notes that the worst-case equivalence of rotation and waterfilling was unproven; we show that it fails.

\paragraph{Rotations.}
Incoherence processing and Hadamard or learned rotations~\citep{chee2023quip,tseng2024quipsharp,ashkboos2024quarot,liu2025spinquant} are standard for removing outliers. QuIP bounds GPTQ's proxy loss under $\mu$-incoherence~\citep{chee2023quip}; our limit is exact for Haar rotations. \citet{keung2026revisiting} show that an optimal \emph{data-dependent} rotation combined with error feedback attains $n\det(\Sigma)^{1/n}$, the lattice bound, via the geometric-mean decomposition; we study \emph{data-independent} rotations.

%% file: sections/discussion.tex
\section{Discussion and Limitations}
\label{sec:discussion}

\paragraph{When is rotation enough?}
\Cref{thm:B} and \cref{prop:outliers,prop:spikes,prop:logscale} together describe when a random rotation can stand in for per-column rate allocation:
\begin{itemize}[leftmargin=*, itemsep=0pt, topsep=1pt]
\item The rotated gap is small when the bulk of the spectrum is narrow on a logarithmic scale and the outliers satisfy $\eta^2\mu\ll D(\nu)$.
\item It grows like $\frac12\log_2\kappa$ when the spectrum is spread on a log scale, or when a fraction $\gtrsim1/\ln\kappa$ of the directions carries the energy.
\end{itemize}
Deciding per layer needs no quantization run, and the rule itself does not use our theorems. For each candidate basis (native order, act-order, and the rotation actually used), compute the exact gap~\eqref{eq:gap} from one Cholesky factorization of the transformed $\Sigma$, add $\Cw$ for the rotation ($\Cw=0$ for a permutation), and keep the cheapest. If even that exceeds the tolerated overhead, use per-column steps. These are rarely needed, but without damping o\_proj inputs reach $0.63$ bit after rotation, and on Llama block~0's q/k/v input the weight-side cost ($1.36$ bit) makes rotation harmful.

\paragraph{Limitations.}
\begin{itemize}[leftmargin=*, itemsep=1pt, topsep=1pt]
\item Our theorems are for Haar rotations, not the randomized Hadamard transforms used in practice, whose worst case remains open. A rotation that makes $\Sigma$ asymptotically free from the coordinate projections, such as a Hadamard transform conjugated by a random signed permutation~\citep{anderson2014asymptotically}, has gap at least $\Delta(\nu)$ in the limit, so its worst case is at least that of \cref{thm:B} (\cref{prop:liberating}). Sign-only transforms are not known to be of this kind (\cref{app:A-liberating}).
\item The gap~\eqref{eq:gap} is a high-rate, entropy-coded quantity, verified down to $3$ bits (conservative by $0.02$--$0.03$ bit at $2$ bits); fixed-rate formats need a different theory. \Cref{prop:S} relies on \cref{ass:H}; the analysis is layerwise.
\end{itemize}

%% file: appendix/proofA.tex
\section{Proof of \Cref{thm:A} and \Cref{prop:props}}
\label{app:A}

Throughout, $\nu$ is a probability measure on $[\epsilon,1]$ and, for $s>0$,
\[
T_\nu(s)=\int\frac{x\,d\nu}{s+x},\qquad A_\nu(s)=\int\frac{d\nu}{s+x},\qquad B_\nu(s)=\int\frac{x\,d\nu}{(s+x)^2}.
\]
Then $T_\nu(s)+sA_\nu(s)=1$ and $T_\nu'(s)=-B_\nu(s)<0$. Write $\bar h_\nu(s):=T_\nu(s)/A_\nu(s)$, so that $h_\nu(t)=\bar h_\nu(s_t)$.

\subsection{Deterministic lemmas}

\begin{lemma}[Well-posedness]
\label{lem:A-wellposed}
$T_\nu$ is a strictly decreasing continuous bijection $(0,\infty)\to(0,1)$, and $s_t\in[\epsilon(1-t),1/t]$.
\end{lemma}
\begin{proof}
Monotonicity and the limits are clear. The bounds follow from $t=T_\nu(s_t)\le\AM/s_t\le1/s_t$ and $1-t=s_tA_\nu(s_t)\le s_t/\epsilon$.
\end{proof}

\begin{lemma}[Monotonicity]
\label{lem:A-monotone}
Let $\rho_s(dx)\propto\nu(dx)/(s+x)$. Then $\frac{d}{ds}\bar h_\nu(s)=-\mathrm{Cov}_{\rho_s}\big(x,\frac1{s+x}\big)\ge0$, with equality iff $\nu$ is a point mass. Moreover $\bar h_\nu(s)\to\HM(\nu)$ as $s\to0$ and $\bar h_\nu(s)\to\AM(\nu)$ as $s\to\infty$. Hence $h_\nu$ is continuous and non-increasing on $[0,1]$, strictly decreasing unless $\nu$ is a point mass, and $\epsilon\le\HM\le h_\nu\le\AM\le1$.
\end{lemma}
\begin{proof}
Note that $\bar h_\nu(s)=\E_{\rho_s}[x]$. Differentiating,
\[
\bar h_\nu'=\E_{\rho_s}[x]\,\E_{\rho_s}\big[\tfrac1{s+x}\big]-\E_{\rho_s}\big[\tfrac{x}{s+x}\big].
\]
Since $x$ is increasing and $1/(s+x)$ strictly decreasing, the covariance is $\le0$, and it is $<0$ unless $\rho_s$ is a point mass (Chebyshev's sum inequality). The limits follow by direct computation. Since $s_t$ decreases in $t$, $h_\nu(t)=\bar h_\nu(s_t)$ is non-increasing in $t$.
\end{proof}

\begin{lemma}[Uniform Lipschitz bound]
\label{lem:A-lipschitz}
For $\delta\in(0,1)$, $t\mapsto1/h_\nu(t)$ is $\frac{2}{\epsilon\delta^2}$-Lipschitz on $[\delta,1]$, uniformly over $\nu\in\cP_\epsilon$.
\end{lemma}
\begin{proof}
We have $1/h_\nu(t)=A_\nu(s_t)/t$ and $ds_t/dt=-1/B_\nu(s_t)$, so
\[
\frac{d}{dt}A_\nu(s_t)=\frac{\int(s+x)^{-2}d\nu}{\int x(s+x)^{-2}d\nu}\in[1,1/\epsilon].
\]
With $A_\nu\le1/\epsilon$ this gives $|\frac{d}{dt}\frac1h|\le\frac1{\epsilon t}+\frac{1}{\epsilon t^2}\le\frac{2}{\epsilon\delta^2}$ on $[\delta,1)$, and continuity extends the bound to $t=1$.
\end{proof}

\begin{lemma}[Continuity in $\nu$]
\label{lem:A-continuity}
If $\nu_n\Rightarrow\nu$ on $[\epsilon,1]$, then $\sup_t|h_{\nu_n}(t)-h_\nu(t)|\to0$. Consequently $D(\nu_n)\to D(\nu)$, $\GM(\nu_n)\to\GM(\nu)$ and $\Delta(\nu_n)\to\Delta(\nu)$.
\end{lemma}
\begin{proof}
For each $s$, $T_{\nu_n}(s)\to T_\nu(s)$. For $t\in(0,1)$ and $s'<s_t(\nu)<s''$ we have $T_\nu(s')>t>T_\nu(s'')$, so $s_t(\nu_n)\in(s',s'')$ for large $n$; hence $h_{\nu_n}(t)\to h_\nu(t)$. At the endpoints, $\AM$ and $\HM$ converge. Monotone functions that converge pointwise to a continuous limit on a compact interval converge uniformly (P\'olya). Finally, $\ln x$ is bounded and continuous on $[\epsilon,1]$.
\end{proof}

\begin{lemma}[Chain rule, \Cref{prop:props}(c)]
\label{lem:A-chain}
$\int_0^1\ln h_\nu(t)\,dt=\int\ln x\,d\nu$. Hence $D(\nu)\ge\GM(\nu)$ by Jensen's inequality, with equality iff $\nu$ is a point mass.
\end{lemma}
\begin{proof}
Let $\Psi(t)=\int\ln(1+x/s_t)\,d\nu+t\ln s_t=\min_{z>0}\big[\int\ln(1+zx)\,d\nu-t\ln z\big]$, attained at $z=1/s_t$. By the envelope theorem $\Psi'(t)=\ln s_t$. Hence $F(t):=\Psi(t)+t\ln t+(1-t)\ln(1-t)$ has $F'(t)=\ln\frac{ts_t}{1-t}=\ln h_\nu(t)$. As $t\to0$, $s_t\sim\AM/t$ and $\Psi\to0$. As $t\to1$, $s_t\to0$ with $1-t\sim s_t/\HM$, so $\Psi(t)=\int\ln(s_t+x)\,d\nu-(1-t)\ln s_t\to\int\ln x\,d\nu$. Therefore $\int_0^1\ln h_\nu=F(1)-F(0)=\int\ln x\,d\nu$. Equality in Jensen's inequality forces $h_\nu$ to be constant, i.e.\ $\nu$ to be a point mass, by \cref{lem:A-monotone}.
\end{proof}

\subsection{Probabilistic tools}

\begin{lemma}[Gaussian representation]
\label{lem:A-gauss}
Let $G\in\R^{n\times k}$ have i.i.d.\ $\mathcal N(0,1)$ entries. Then $G(G^\top G)^{-1/2}$ has the law of the first $k$ columns of a Haar orthogonal matrix.
\end{lemma}
\begin{proof}
The law is invariant under left multiplication by $O(n)$. The Stiefel manifold is a homogeneous space of $O(n)$, and an invariant probability measure on it is unique. (Equivalently, use the Gram--Schmidt construction $G_kR_k^{-1}$, which differs by a $k\times k$ orthogonal factor and leaves all traces used below unchanged; see \citet{meckes2019random}.)
\end{proof}

\begin{lemma}[Spherical representation]
\label{lem:A-sphere}
Let $M=\tilde\Sigma^{(k)}$ be the leading $k\times k$ block of $\tilde\Sigma=V^\top\Lambda V$, with eigenvalues $\gamma_1,\dots,\gamma_k$, and let $m_k=\frac1k\tr M^{-1}$. Then $\pi_k=1/(M^{-1})_{kk}$, every $\gamma_i\ge\lambda_{\min}(\Lambda)$, and
\[
\big((M^{-1})_{kk},\,m_k\big)\overset{d}{=}\Big(\sum_iw_i^2/\gamma_i,\ \tfrac1k\sum_i1/\gamma_i\Big),
\]
with $w$ uniform on $S^{k-1}$ and independent of $\gamma$. This holds for every diagonal $\Lambda$.
\end{lemma}
\begin{proof}
The first identity is the Schur complement formula, and the second follows from Cauchy interlacing. For $O\in O(k)$, $V\diag(O,I_{n-k})$ is again Haar, so $M\overset{d}{=}O^\top MO$. Write $M=U\Gamma U^\top$ and take an independent Haar $O'\in O(k)$. Then $W=O'^\top U$ is Haar given $V$, hence independent of $\Gamma$, and $w=W^\top e_k$ is uniform on $S^{k-1}$.
\end{proof}

\begin{lemma}[\citealp{vershynin2012introduction}, Cor.~5.35]
\label{lem:A-svd}
For an $N\times k$ standard Gaussian matrix $A$ and $r\ge0$, with probability at least $1-2e^{-r^2/2}$,
\[
\sqrt N-\sqrt k-r\le s_{\min}(A)\le s_{\max}(A)\le\sqrt N+\sqrt k+r.
\]
\end{lemma}

\begin{lemma}[Hanson--Wright, Gaussian case of \citealp{rudelson2013hanson}, Thm.~1.1]
\label{lem:A-hw}
There is an absolute constant $c>0$ such that, for $z\sim\mathcal N(0,I_k)$ and a fixed matrix $M$,
\[
\Prob\big(|z^\top Mz-\tr M|>r\big)\le2\exp\big[-c\min(r^2/\|M\|_F^2,\,r/\|M\|)\big].
\]
\end{lemma}

\subsection{Proof of \Cref{thm:A}}

Let $\Theta_k=\tr(\tilde\Sigma^{(k)})^{-1}=k\,m_k$. Every ``almost surely'' below follows from Borel--Cantelli: for each $r>0$, the probability of the bad event is summable in $n$. The coupling of $V_n$ across different $n$ is irrelevant.

\paragraph{Step 1: $\pi_k\approx1/m_k$ uniformly for $k\ge\delta n$.}
By \cref{lem:A-sphere}, conditionally on $\gamma$, $(M^{-1})_{kk}\overset{d}{=}z^\top\mathcal Az/z^\top z$ with $z\sim\mathcal N(0,I_k)$ and $\mathcal A=\diag(1/\gamma_i)$, where $1/\gamma_i\in[1,1/\epsilon]$. Let $X=z^\top(\mathcal A-m_kI)z$. Then $\E X=0$, $\|\mathcal A-m_kI\|\le1/\epsilon$ and $\|\mathcal A-m_kI\|_F^2\le k/\epsilon^2$. By \cref{lem:A-hw},
\[
\Prob(|X|>rk/2)\le2\exp[-ck\min(r^2\epsilon^2/4,r\epsilon/2)],\qquad \Prob(z^\top z<k/2)\le2e^{-ck/4}.
\]
Outside these events $|(M^{-1})_{kk}-m_k|\le r$. Since $(M^{-1})_{kk},m_k\ge1$, also $|\pi_k-1/m_k|\le|(M^{-1})_{kk}-m_k|$. A union bound over $k\in[\delta n,n]$ gives a failure probability $\le4n\exp[-c\delta n\min(r^2\epsilon^2/4,r\epsilon/2,1/4)]$, which is summable.

\paragraph{Step 2: for fixed $\tau\in(0,1)$ and $k=\lceil\tau n\rceil$, $m_k-1/h_{\nu_n}(k/n)\to0$ a.s.}
By \cref{lem:A-gauss}, $\tilde\Sigma^{(k)}\overset{d}{=}S^{-1/2}BS^{-1/2}$ with $S=G^\top G$ and $B=\sum_{j=1}^n\lambda_jg_jg_j^\top$, where $g_j^\top$ are the rows of $G$. Hence $\Theta_k\overset{d}{=}\tr(B^{-1}S)=\sum_jg_j^\top B^{-1}g_j$. Let $B_j=B-\lambda_jg_jg_j^\top$ and $q_j=g_j^\top B_j^{-1}g_j$. Sherman--Morrison gives $g_j^\top B^{-1}g_j=q_j/(1+\lambda_jq_j)$, hence the two exact identities
\[
\text{(I1)}\ \ \Theta_k=\sum_j\frac{q_j}{1+\lambda_jq_j},\qquad\text{(I2)}\ \ k=\tr(B^{-1}B)=\sum_j\frac{\lambda_jq_j}{1+\lambda_jq_j}.
\]

\emph{Good event.} Let $\tau_1=(1+\tau)/2$, so $k\le\tau_1n$ for large $n$. Apply \cref{lem:A-svd} to $G_{-j}$ (row $j$ removed) with $r=\beta\sqrt n$, $\beta=(1-\sqrt{\tau_1})/4$. Then $\lambda_{\min}(G_{-j}^\top G_{-j})\ge c_0n$ with $c_0=(1-\sqrt{\tau_1})^2/4$, for all $j$, with failure probability $\le2ne^{-\beta^2n/2}$. On this event $B_j\succeq\epsilon G_{-j}^\top G_{-j}$ and $B\succeq B_j$, so $\|B_j^{-1}\|,\|B^{-1}\|\le(\epsilon c_0n)^{-1}$.

\emph{Concentration of $q_j$.} Given $G_{-j}$, $g_j\sim\mathcal N(0,I_k)$ is independent of $B_j$, and $\|B_j^{-1}\|_F^2\le k\|B_j^{-1}\|^2\le(\epsilon^2c_0^2n)^{-1}$. By \cref{lem:A-hw},
\[
\Prob\big(|q_j-\tr B_j^{-1}|>r\big)\le2\exp[-c\min(r^2\epsilon^2c_0^2n,\ r\epsilon c_0n)]
\]
plus the failure probability of the good event. A rank-one update gives
\[
0\le\tr B_j^{-1}-\tr B^{-1}=\frac{\lambda_jg_j^\top B_j^{-2}g_j}{1+\lambda_jq_j}\le\frac{1}{\epsilon c_0n}.
\]
Let $\vartheta=\tr B^{-1}$ and $\rho_n=\max_j|q_j-\vartheta|$. Then $\Prob(\rho_n>r+\frac1{\epsilon c_0n})$ is summable in $n$.

\emph{A caveat.} The Gaussian representation holds only in distribution, so we cannot conclude $\rho_n\to0$ a.s.\ on the probability space of $V_n$ directly. Instead we bound $|m_k-1/h_{\nu_n}(k/n)|$ by a deterministic function of $\rho_n$. This yields a summable bound on $\Prob(|m_k-1/h_{\nu_n}(k/n)|>r)$, a probability that depends only on the law of $m_k$, and we then apply Borel--Cantelli on the space of $V_n$.

\emph{Solving for $s$.} Let $t_n=k/n$, $s=1/\vartheta$ and $s^*=s_{t_n}(\nu_n)$. The maps $q\mapsto\frac{\lambda q}{1+\lambda q}$ and $q\mapsto\frac{q}{1+\lambda q}$ are $1$-Lipschitz for $\lambda\le1$. So (I2) gives $|t_n-T_{\nu_n}(s)|\le\rho_n$ and (I1) gives $|\Theta_k/n-A_{\nu_n}(s)|\le\rho_n$, using $\frac{\vartheta}{1+\lambda\vartheta}=\frac{1}{s+\lambda}$. Next, $\vartheta\le k\|B^{-1}\|\le(\epsilon c_0)^{-1}$, and $\vartheta\ge T_{\nu_n}(s)\ge t_n-\rho_n\ge\tau/2$ for large $n$. Thus $s\in[\epsilon c_0,2/\tau]$, and $s^*\in[\epsilon(1-\tau_1),2/\tau]$ by \cref{lem:A-wellposed}. Let $b=2/\tau$. On this range $-T_{\nu_n}'\ge\epsilon/(b+1)^2$, so $|s-s^*|\le\rho_n(b+1)^2/\epsilon$. Since $A_{\nu_n}$ is $\epsilon^{-2}$-Lipschitz and $A_{\nu_n}(s^*)=t_n/h_{\nu_n}(t_n)$,
\[
\Big|m_k-\frac1{h_{\nu_n}(t_n)}\Big|=\frac{1}{t_n}\Big|\frac{\Theta_k}{n}-A_{\nu_n}(s^*)\Big|\le\frac{\rho_n}{t_n}\Big(1+\frac{(b+1)^2}{\epsilon^3}\Big)
\]
on the good event when $\rho_n\le\tau/4$.

\paragraph{Step 3: uniformity over $k\in[\delta n,n]$.}
The eigenvalues of $\tilde\Sigma^{(k+1)}$ and $\tilde\Sigma^{(k)}$ interlace in $[\epsilon,1]$, so $\Theta_{k+1}-\Theta_k\in[1,1/\epsilon]$. Hence
\[
m_{k+1}-m_k=\frac{\Theta_{k+1}-\Theta_k}{k+1}-\frac{\Theta_k}{k(k+1)}
\]
has both terms in $[0,\frac1{\epsilon(k+1)}]$, and $|m_{k+1}-m_k|\le\frac1{\epsilon\delta n}$ for $k\ge\delta n$. Take a grid $\delta=\tau_0<\dots<\tau_M=1$ with mesh $\le\eta$. Use Step~2 at the interior points and the exact identity $m_n=1/\HM(\nu_n)=1/h_{\nu_n}(1)$ at $\tau_M=1$. With \cref{lem:A-lipschitz},
\[
\limsup_n\max_{k\ge\delta n}\Big|m_k-\frac1{h_{\nu_n}(k/n)}\Big|\le\frac{\eta}{\epsilon\delta}+\frac{2\eta}{\epsilon\delta^2}\quad\text{a.s.}
\]
Letting $\eta=1/j\to0$ along countably many events, and using $h\le1\le m_k$ (so that $|1/m_k-h|\le|m_k-1/h|$), Step~1 yields (i).

\paragraph{Step 4: (ii) and (iii).}
Both $\pi_k\le\tilde\Sigma_{kk}\le1$ and $h_{\nu_n}\le1$, so the indices $k<\delta n$ contribute at most $\delta$ to each side. The Riemann-sum error of the monotone function $h_{\nu_n}$ is at most $2/n$. Hence $\limsup|\frac1n\sum\pi_k-D(\nu_n)|\le2\delta$ for every $\delta$, which proves (ii). For (iii), $\GM(\nu_n)=(\det\Lambda_n)^{1/n}$ is rotation invariant, and $\frac1n\sum\pi_k,D(\nu_n)\ge\epsilon$ gives $|\ln\frac1n\sum\pi_k-\ln D(\nu_n)|\le\epsilon^{-1}|\frac1n\sum\pi_k-D(\nu_n)|$.

\paragraph{Explicit probability bound.}
Tracking the probabilities through Steps~1--3, with grid mesh $\eta=r\epsilon\delta^2/9$ and accuracy $r/3$ in each step, gives
\[
\Prob\Big(\max_{\lceil\delta n\rceil\le k\le n}|\pi_k-h_{\nu_n}(k/n)|>r\Big)\le Cne^{-cn}
\]
with $C,c$ depending only on $(\epsilon,\delta,r)$, uniformly over spectra in $[\epsilon,1]$. The bounds for (ii) and (iii) follow by taking $\delta=r/8$, and replacing $r$ by $\epsilon r$ for (iii). \hfill$\square$

\paragraph{Order convention.}
GPTQ's effective pivots are conditional variances given the \emph{later} coordinates (reverse Cholesky). Under a Haar rotation the law of $\tilde\Sigma$ is invariant under coordinate permutations, so the forward and reverse pivot sequences have the same law and \cref{thm:A} applies to both.

\subsection{Free-probability form and the annealed approximation}
\label{app:A-free}

\paragraph{$S$-transform.}
With $\psi_\nu(w)=\int\frac{wx}{1-wx}d\nu$ and $S_\nu(z)=(1+z^{-1})\psi_\nu^{-1}(z)$, taking $w=-1/s_t$ gives $\psi_\nu(w)=-T_\nu(s_t)=-t$. Hence
\[
S_\nu(-t)=\big(1-\tfrac1t\big)\big(-\tfrac1{s_t}\big)=\frac{1-t}{ts_t}=\frac1{h_\nu(t)}.
\]
Here $\psi_\nu$ is real analytic and strictly monotone on $(-\infty,0)$ with range $(-1,0)$, so the real inverse agrees with the analytic inverse near the origin.

\paragraph{Free compression.}
Let $\nu_t:=D_t(\nu^{\boxplus1/t})$ be the compression of $\nu$ by a free projection of trace $t$, with $R_{\nu_t}(w)=R_\nu(tw)$~\citep{nica1996multiplication,nica2006lectures}. Let $g=G_{\nu_t}(0)=-\int x^{-1}d\nu_t$. The relation $0=R_{\nu_t}(g)+1/g$ gives $R_\nu(tg)=-1/g$. With $u=tg$ this becomes $G_\nu^{-1}(u)=(1-t)/u$, i.e.\ $zG_\nu(z)=1-t$ at $z=(1-t)/u=-s$, which is $T_\nu(s)=t$. Therefore $\HM(\nu_t)=-1/g=ts/(1-t)=h_\nu(t)$, which is \cref{prop:props}(a).

\paragraph{The annealed approximation.}
\citet[Eq.~(33)]{ordentlich2026hrqmm2} replace the pivots by ratios of expected leading minors,
\[
\pi_k^{\rm ann}:=\frac{k}{n-k+1}\,\frac{e_k(\lambda)}{e_{k-1}(\lambda)} ,
\]
where $e_k$ is the $k$-th elementary symmetric polynomial. The next result shows that this approximation is uniformly accurate. The bound is deterministic and has an explicit rate.

\begin{proposition}[Annealed approximation]
\label{prop:annealed}
Fix $\epsilon\in(0,1)$ and $\delta\in(0,\frac12)$. For all $n\ge2$ and all $\lambda_1,\dots,\lambda_n\in[\epsilon,1]$ with empirical measure $\nu_n$,
\[
\max_{\delta n\le k\le(1-\delta)n}\Big|\ln\pi_k^{\rm ann}-\ln h_{\nu_n}(k/n)\Big|\le\Big(\frac{16}{\epsilon\delta^2}+4\Big)\sqrt{\frac{\ln(n+1)}{n}} .
\]
Consequently, by \cref{thm:A}, $\max_{\delta n\le k\le(1-\delta)n}|\pi_k-\pi_k^{\rm ann}|\to0$ almost surely after a Haar rotation: the quenched pivots and the annealed approximation have the same limit.
\end{proposition}

\begin{proof}
Write $E_k=e_k(\lambda)/\binom nk$ and $a_k=\ln E_k$. Since $\binom nk/\binom n{k-1}=\frac{n-k+1}{k}$, we have the exact identity
\[
\pi_k^{\rm ann}=E_k/E_{k-1}=\exp(a_k-a_{k-1}).
\]

\emph{Step 1: $a$ is concave (Newton's inequalities).} We show $E_k^2\ge E_{k-1}E_{k+1}$. The polynomial $P(x)=\prod_j(1+\lambda_jx)=\sum_m\binom nmE_mx^m$ has only real negative roots, and so do all its derivatives (Rolle).
\begin{itemize}[leftmargin=*, itemsep=0pt, topsep=1pt]
\item Differentiate $k-1$ times.
\item Reverse the result, i.e.\ form $x^{n-k+1}P^{(k-1)}(1/x)$; this keeps real-rootedness because no root is $0$.
\item Differentiate $n-k-1$ more times.
\end{itemize}
A direct computation shows that what remains is the quadratic $\frac{n!}2\,(E_{k+1}+2E_kx+E_{k-1}x^2)$. It has real roots, so its discriminant is nonnegative, which is the claim. Since $\binom nk$ is also log-concave in $k$, $e_k$ is log-concave too.

\emph{Step 2: exponential tilting.} For $z>0$, let $S_z=\sum_jB_j$ with independent $B_j\sim\mathrm{Bernoulli}\big(\frac{z\lambda_j}{1+z\lambda_j}\big)$. Expanding the product gives
\[
\Prob(S_z=k)=\frac{z^ke_k}{\prod_j(1+z\lambda_j)}.
\]
Write $G(z)=\sum_j\ln(1+z\lambda_j)-k\ln z$, so that $\ln e_k=G(z)+\ln\Prob(S_z=k)$ for every $z$. We have $G'(z)=(\E S_z-k)/z$. So $G$ is minimized, and $\Prob(S_z=k)=e_k e^{-G(z)}$ is maximized, at the $z_k$ with $\E S_{z_k}=k$, namely $z_k=1/s_{k/n}(\nu_n)$.

\emph{Step 3: $\Prob(S_{z_k}=k)\ge\frac1{n+1}$.} Let $d_j=\ln e_j-\ln e_{j-1}$, which is non-increasing by Step~1. Pick $\theta$ with $-\theta\in[d_{k+1},d_k]$. Then $j\mapsto j\theta+\ln e_j$ is maximized at $j=k$, i.e.\ $k$ is a mode of $S_{e^\theta}$. Hence $\Prob(S_{e^\theta}=k)\ge\frac1{n+1}$, and by Step~2 also $\Prob(S_{z_k}=k)\ge\frac1{n+1}$.

\emph{Step 4: the limit profile.} Let $\Psi_n(t)=\min_{z>0}\big[\int\ln(1+zx)\,d\nu_n-t\ln z\big]$, so that $G(z_k)=n\Psi_n(k/n)$. Steps~2--3 give
\[
|\ln e_k-n\Psi_n(k/n)|\le\ln(n+1).
\]
Applying the same steps to $\lambda\equiv1$ gives $|\ln\binom nk-nH(k/n)|\le\ln(n+1)$, where $H(t)=-t\ln t-(1-t)\ln(1-t)$. Therefore
\[
a_k=nF_n(k/n)+r_k,\qquad |r_k|\le2\ln(n+1),\qquad F_n(t):=\Psi_n(t)+t\ln t+(1-t)\ln(1-t).
\]
By the proof of \cref{lem:A-chain}, $F_n'(t)=\ln h_{\nu_n}(t)$. By \cref{lem:A-lipschitz} and $1/h\ge1$, $\ln h_{\nu_n}$ is $L_\delta$-Lipschitz on $[\delta/2,1]$, with $L_\delta=8/(\epsilon\delta^2)$ uniform in $\nu_n$.

\emph{Step 5: chords.} Take $1\le m\le\delta n/2-1$. By concavity,
\[
\frac{a_{k+m}-a_k}{m}\le a_k-a_{k-1}\le\frac{a_{k-1}-a_{k-1-m}}{m}.
\]
By Step~4 and the mean-value theorem, each chord equals $\ln h_{\nu_n}(\xi)+O(\ln(n+1)/m)$ for some $\xi$ within $(m+1)/n$ of $k/n$. Hence
\[
\big|(a_k-a_{k-1})-\ln h_{\nu_n}(k/n)\big|\le\frac{L_\delta(m+1)}{n}+\frac{4\ln(n+1)}{m}.
\]
Let $\rho_n=\sqrt{\ln(n+1)/n}$ and $m=\lceil\sqrt{n\ln(n+1)}\rceil$.
\begin{itemize}[leftmargin=*, itemsep=0pt, topsep=1pt]
\item If $\sqrt{n\ln(n+1)}+2\le\delta n/2$, then $m+1\le\delta n/2$ and $n>4/\delta>8$, so $2/n\le\rho_n$. The right-hand side of the last display is at most $L_\delta(\rho_n+2/n)+4\rho_n\le(2L_\delta+4)\rho_n=(\frac{16}{\epsilon\delta^2}+4)\rho_n$.
\item Otherwise $\rho_n>\delta/4$: for $n\ge3$ because then $\sqrt{n\ln(n+1)}\ge2$, so $\delta n/2<2\sqrt{n\ln(n+1)}$; for $n=2$ because $\rho_2>0.7$. Both $a_k-a_{k-1}$ and $\ln h_{\nu_n}(k/n)$ lie in $[\ln\epsilon,0]$, the former because it is non-increasing from $\ln\AM(\nu_n)$ to $\ln\HM(\nu_n)$. So the error is at most $\ln\frac1\epsilon<\frac4\delta\ln\frac1\epsilon\,\rho_n\le\frac{16}{\epsilon\delta^2}\rho_n$, using $\ln\frac1\epsilon<\frac1\epsilon$ and $\delta<4$.
\end{itemize}
\end{proof}

\paragraph{Rate and constant.}
The constant grows like $\kappa/\delta^2$ with $\kappa=1/\epsilon$. At the condition numbers of real layers ($10^3$--$10^5$, \cref{app:e3}) and $n\le14336$ the bound exceeds $\ln\kappa$ and says nothing; its role is to make the limit rigorous and uniform. The rate $\sqrt{\ln n/n}$ comes from the chord argument and is not sharp. Numerically the error decays like $1/n$: for the log-uniform spectrum with $\epsilon=10^{-2}$ and for $\nu_{0.2}$ with $\epsilon=10^{-2}$, $n$ times the maximal error over $0.1n\le k\le0.9n$ stays near $1.3$ and between $7.2$ and $7.6$, respectively, for $n=50$--$400$.

\paragraph{\Cref{prop:props}(b), (d), (e).}
Part (b) is \cref{lem:A-monotone}. In (d), $D\le\AM$ holds because $h_\nu\le h_\nu(0)=\AM$, strictly unless $\nu$ is a point mass, and $D\ge\GM$ is \cref{lem:A-chain}. For (e), $x\mapsto x/(s+x)$ is increasing, so $T_\nu\le T_{\nu'}$ pointwise, hence $s_t(\nu)\le s_t(\nu')$ and $h_\nu\le h_{\nu'}$.

\subsection{Beyond Haar rotations}
\label{app:A-liberating}

For a random orthogonal $R_n$ let $\pi_k$ be the pivots~\eqref{eq:pivots} of $\Sigma_n^R:=R_n^\top\Sigma_nR_n$, and let $F_n(t):=\frac1n\ln\det(\Sigma_n^R)^{(\lfloor tn\rfloor)}=\frac1n\sum_{k\le tn}\ln\pi_k$. Consider the condition
\begin{itemize}[leftmargin=2.4em, itemsep=0pt, topsep=1pt]
\item[(L)] for every $t\in(0,1)$, the empirical spectral measure of $(\Sigma_n^R)^{(\lfloor tn\rfloor)}$ converges weakly to the free compression $\nu_t$ of \cref{app:A-free}, almost surely.
\end{itemize}
It says that $\Sigma_n^R$ is asymptotically free from the coordinate projections. Haar rotations satisfy it, and so do the \emph{fake Haar} rotations of \citet{anderson2014asymptotically}.

\begin{proposition}[Liberating rotations are no better than Haar]
\label{prop:liberating}
Let $\Sigma_n$ be deterministic with spectrum in $[\epsilon,1]$ and $\nu_n\Rightarrow\nu$.
\begin{enumerate}[label=(\roman*), leftmargin=*, itemsep=0pt, topsep=1pt]
\item If (L) holds, then almost surely $\liminf_n\frac1n\sum_k\pi_k\ge D(\nu)$ and $\liminf_n\Delta(\Sigma_n^R)\ge\Delta(\nu)$.
\item (L) holds for $R_n=W_n^\top H_nW_n/\sqrt n$, where $W_n$ is a uniformly random signed permutation matrix and $H_n$ a deterministic Hadamard matrix ($n$ runs through dimensions for which one exists).
\item For such rotations, taking $\Sigma_n$ diagonal at the quantiles of $\nu$ gives $\liminf_n\Delta(\Sigma_n^R)\ge\Delta(\nu)$ for every $\nu\in\cP_\epsilon$. So the worst case is at least $\Delta^*(\epsilon)$ of \cref{thm:B}.
\end{enumerate}
\end{proposition}

\begin{proof}
\emph{Block AM--GM.} Fix $m\ge2$, $\delta=1/m$, $k_i=\lfloor i\delta n\rfloor$ and $m_i=k_i-k_{i-1}$. Within each block of pivots the arithmetic mean is at least the geometric mean, so
\[
\frac1n\sum_{k=1}^n\pi_k\ \ge\ \sum_{i=1}^m\frac{m_i}n\exp\Big(\frac n{m_i}\big(F_n(i\delta)-F_n((i-1)\delta)\big)\Big).
\]

\emph{The limit profile.} Let $\Phi(t):=t\int\ln x\,d\nu_t(x)$. By interlacing, the eigenvalues of every leading block lie in $[\epsilon,1]$, where $\ln$ is bounded and continuous. Hence (L) gives $F_n(t)\to\Phi(t)$ almost surely. We claim that $\Phi(t)=\int_0^t\ln h_\nu(s)\,ds$. Haar rotations satisfy the hypotheses of \citet[Thm.~2.8]{anderson2014asymptotically} (their \S2.9.5). By their Proposition~2.6, the expected empirical spectral measure of the leading block then converges to $\nu_t$, so $\E F_n(t)\to\Phi(t)$. On the other hand, \cref{thm:A}(i) and \cref{lem:A-continuity} give $F_n(t)\to\int_0^t\ln h_\nu$ almost surely. Since $|\ln\pi_k|\le\ln\frac1\epsilon$, they also give it in expectation.

\emph{Proof of (i).} $F_n(0)=0$ and $F_n(1)=\frac1n\ln\det\Sigma_n$ need no assumption. Letting $n\to\infty$ in the block bound gives, almost surely,
\[
\liminf_n\frac1n\sum_k\pi_k\ \ge\ \sum_{i=1}^m\delta\exp\Big(\frac1\delta\int_{(i-1)\delta}^{i\delta}\ln h_\nu\Big)=\sum_{i=1}^m\delta\,h_\nu(\xi_i)
\]
for some $\xi_i\in((i-1)\delta,i\delta)$, by the mean-value theorem. This is a Riemann sum of the continuous function $h_\nu$ (\cref{lem:A-monotone}), and it tends to $D(\nu)$ as $\delta\to0$. Since $\GM(\nu_n)\to\GM(\nu)$, the claim on $\Delta$ follows.

\emph{Proof of (ii).} \citet{anderson2014asymptotically} state their results for unitary matrices. Their signed permutation matrices have entries $\pm1$ (their \S2.7), a Hadamard matrix in their sense is real with $H_n/\sqrt n$ orthogonal (\S3.1), and a real orthogonal matrix is unitary, so the results below apply as stated. Note that $U_n:=R_n^\top$ has the same form, since $H_n^\top$ is a Hadamard matrix.
\begin{itemize}[leftmargin=*, itemsep=0pt, topsep=1pt]
\item By \citet[Cor.~3.5]{anderson2014asymptotically}, the leading-coordinate projection $P=P_{\lfloor tn\rfloor}$ and $U_n\Sigma_nU_n^\top$ are asymptotically free with respect to $\frac1n\E\tr$.
\item The spectrum of $PU_n\Sigma_nU_n^\top P$ is that of the leading block plus $n-\lfloor tn\rfloor$ zeros. Hence its expected empirical spectral measure converges to $(1-t)\delta_0+t\nu_t$.
\item $W_0U_nW_0^\top$ has the same law as $U_n$ for every signed permutation matrix $W_0$. So \citet[Thm.~7.1]{anderson2014asymptotically} applies: at every $x$, the empirical distribution function $F$ of this matrix satisfies $\Prob(|F(x)-\E F(x)|>s)\le2\exp(-ns^2/(128(\log n+c)))$, where $c$ does not depend on $n$ or on the matrices.
\item By Borel--Cantelli over rational $x$, the convergence holds almost surely.
\end{itemize}
Part (iii) follows from (i) and (ii).
\end{proof}

\paragraph{Scope.} Haar rotations attain equality (\cref{thm:A}). So among rotations that make $\Sigma$ asymptotically free from the coordinate structure, Haar is asymptotically optimal for every spectrum, and doing better requires exploiting structure.

The randomized Hadamard transforms used in practice, $DH_n$ and $D_1H_nD_2H_n$ with random signs only, are not covered. Whether the random permutation can be dropped is left open by \citet[Question~3.3]{anderson2014asymptotically}, and we found no later result that settles it. For approximate message passing, random signs alone do give the same dynamics as a Haar matrix \citep{dudeja2023universality}, but that result concerns the algorithm, not spectra of compressions, and its authors note that it does not follow from free probability (their Remark~7). For one round the permutation cannot be dropped in general. On axis-aligned spectra one round gives gaps below $\Delta(\nu)$ that do not approach it (\cref{tab:e1}). By (i), this is incompatible with (L) if it persists as $n\to\infty$. For two rounds we have no such evidence. Their gaps on o\_proj inputs are below $\Delta(\nu)$, but inserting a random permutation brings them above it on every layer. On block-diagonal covariances built from the attention heads, the deficit also decays as $n$ grows (\cref{app:e3}).

Part (ii) also holds with two independent uniformly random signed permutations, $R_n=W_n'^\top H_nW_n/\sqrt n$. Indeed, the pair $(W_n,\,H_n^\top W_n'/\sqrt n)$ satisfies the hypotheses of \citet[Thm.~2.8]{anderson2014asymptotically}: $(W_nW_0,W_n'W_0)$ has the same law as $(W_n,W_n')$ for every signed permutation $W_0$, and $W_n^\top H_n^\top W_n'$ has entries $\pm1$. Their Theorem~7.1 applies to $R_n^\top$ as before.

The matching upper bound $\Delta(\Sigma_n^R)\le\Delta(\nu)+o(1)$ for fake Haar rotations is open. The independent version shows where the difficulty lies. There $\Sigma_n^R=W_n^\top M_nW_n$ with $M_n$ independent of $W_n$, so the column order is uniformly random: given $M_n$ and the set of the first $k$ coordinates, the $k$-th is uniform among them, and $\E[1/\pi_k]$ is the normalized trace of the inverse of the leading $k\times k$ block, which (L) controls. But $1/\pi_k$ is a single diagonal entry of that inverse. An upper bound needs these entries to concentrate around their mean, uniformly in $k$, so that the pivots within a block are nearly equal. For Haar rotations this is Step~1 of the proof of \cref{thm:A}, which uses the rotation invariance of the leading block (\cref{lem:A-sphere}). Second-order freeness, which describes fluctuations of traces, does not supply this concentration. Until this is resolved, the positive results of \cref{sec:typical} are proved for Haar rotations only.

%% file: appendix/proofB.tex
\section{Proof of \Cref{thm:B} and \Cref{thm:Bprime}}
\label{app:B}

Let $L=\ln(1/\epsilon)$, $J(\nu)=\ln D(\nu)-\int\ln x\,d\nu$ (so $\Delta=J/(2\ln2)$) and $\nu_q=(1-q)\delta_\epsilon+q\delta_1$.

\subsection{Existence}
$J$ is continuous on the weakly compact set $\cP_\epsilon$ (\cref{lem:A-continuity}), so maximizers exist. By \cref{lem:A-chain}, $J\ge0$ with equality only at point masses, and $J(\nu_{1/2})>0$, so no maximizer is a point mass.

\subsection{First variation}

\begin{lemma}[Gateaux derivative]
\label{lem:B-gateaux}
For $\nu,\mu\in\cP_\epsilon$ and $\nu_\eta=(1-\eta)\nu+\eta\mu$,
\[
\frac{d}{d\eta}D(\nu_\eta)\Big|_{\eta=0^+}=\int\Phi_\nu\,d(\mu-\nu),\qquad
\Phi_\nu(y)=-\int_0^1\frac{\bar h_\nu(s)}{s+y}ds+\int_1^\infty\frac{\bar h_\nu(s)\,y}{s(s+y)}ds .
\]
\end{lemma}
\begin{proof}
\emph{(a) Differentiability.} Fix $t\in(0,1)$ and let $F(s,\eta)=(1-\eta)T_\nu(s)+\eta T_\mu(s)-t$. $F$ is affine in $\eta$, smooth in $s$, and $\partial_sF=-B_{\nu_\eta}<0$ for $\eta\in[0,1]$, hence on an open neighbourhood of $[0,1]$. The implicit function theorem and uniqueness of the root give $\partial_\eta s_t=\frac{T_\mu(s)-T_\nu(s)}{B_{\nu_\eta}(s)}\big|_{s=s_t(\eta)}$.

\emph{(b) Uniform bound.} For any $\rho\in\cP_\epsilon$ and $t=T_\rho(s)$ we have three facts:
\begin{itemize}[leftmargin=*, itemsep=0pt, topsep=1pt]
\item $B_\rho(s)\ge\epsilon t(1-t)/s$. To see this, write $B_\rho=\frac1s\int\phi(1-\phi)\,d\rho$ with $\phi=x/(s+x)$, and note $1-\phi(x)\ge\epsilon(1-t)$.
\item $|T_\mu(s)-T_\nu(s)|\le\min(s/\epsilon,1/s)$.
\item $1-t\ge s/(s+1)$.
\end{itemize}
Hence $\big|\frac{t}{1-t}\partial_\eta s_t\big|\le\frac{\min(s^2/\epsilon,1)}{\epsilon(1-t)^2}\le\frac4{\epsilon^2}$, uniformly in $t\in(0,1)$ and $\eta\in[0,1]$.

\emph{(c) Differentiation.} $D(\nu_\eta)=\int_0^1\frac{ts_t(\eta)}{1-t}dt$. By dominated convergence,
\[
\frac{d}{d\eta}D\big|_{0^+}=\int_0^1\frac{t}{1-t}\frac{T_\mu(s_t)-T_\nu(s_t)}{B_\nu(s_t)}dt .
\]

\emph{(d) Change of variables.} Substituting $t=T_\nu(s)$ and using $\frac{t}{1-t}=\bar h_\nu(s)/s$ gives $\int_0^\infty\frac{\bar h_\nu(s)}{s}[T_\mu(s)-T_\nu(s)]\,ds$.

\emph{(e) Fubini.} Write $T_\mu(s)-T_\nu(s)=\int k(s,y)\,d(\mu-\nu)(y)$ with $k(s,y)=\frac{y}{s+y}-\mathbf1\{s<1\}$. The integrand is bounded by $1/\epsilon$ for $s<1$ and by $1/s^2$ for $s\ge1$, so Fubini applies.
\end{proof}

Consequently $\frac{d}{d\eta}J(\nu_\eta)|_{0^+}=\int\psi_\nu\,d(\mu-\nu)$ with $\psi_\nu(y)=\Phi_\nu(y)/D(\nu)-\ln y$.

\subsection{Log-convexity (\Cref{lem:logconvex})}
\begin{proof}[Proof of \cref{lem:logconvex}]
Differentiating under the integral (dominated convergence on compacts),
\[
\Phi_\nu'(y)=\int_0^\infty\frac{\bar h_\nu(s)}{(s+y)^2}ds .
\]
With $y=e^u$ and $s=e^v$,
\[
y\Phi_\nu'(y)=\int\bar h_\nu(e^v)\frac{e^{u+v}}{(e^u+e^v)^2}dv=\int H(v)\,\ell(u-v)\,dv,\qquad H(v)=\bar h_\nu(e^v).
\]
By \cref{lem:A-monotone}, $H$ is strictly increasing when $\nu$ is not a point mass. Since $\ell>0$ and $\int\ell=1$, the convolution $H*\ell$ is strictly increasing. Hence $u\mapsto\Phi_\nu(e^u)$ is strictly convex, and so is $\psi_\nu\circ\exp$, which differs from it by a linear function.
\end{proof}

\subsection{Extremal structure}
Let $\nu^*$ maximize $J$ and set $\psi=\psi_{\nu^*}$. For every $\mu$, $0\ge\int\psi\,d(\mu-\nu^*)$. Taking $\mu=\delta_y$ shows $\psi(y)\le c:=\int\psi\,d\nu^*$ on $[\epsilon,1]$. Hence $c-\psi\ge0$ is continuous with $\int(c-\psi)\,d\nu^*=0$, so $\psi=c=\max\psi$ on $\supp\nu^*$. By \cref{lem:logconvex}, $\psi\circ\exp$ is strictly convex on $[\ln\epsilon,0]$, so its maximum is attained only at the endpoints. Therefore $\supp\nu^*\subset\{\epsilon,1\}$, and both atoms are present because point masses have $J=0$. This proves \cref{thm:B}(i).

\subsection{Two-point bounds and asymptotics}

\begin{lemma}
\label{lem:B-twopoint}
Let $g(r)=r+(1-r)\ln(1-r)=\sum_{m\ge2}\frac{r^m}{m(m-1)}$. For $q\in[0,1]$,
\[
(1+\epsilon)\,g\Big(\frac{q}{1+\epsilon}\Big)\le D(\nu_q)\le g\big(\min(q+\epsilon,1)\big)+\frac\pi2\sqrt\epsilon ,
\qquad\text{and}\qquad \frac{r^2}{2}\le g(r)\le\frac{r^2}{2(1-r)}.
\]
\end{lemma}
\begin{proof}
With $p=1-q$, the equation $t=\frac{p\epsilon}{s+\epsilon}+\frac{q}{s+1}$ becomes $ts^2+bs-\epsilon(1-t)=0$ with $b=t(1+\epsilon)-(q+p\epsilon)$. Hence $ts=\frac12(-b+\sqrt{b^2+4t\epsilon(1-t)})\in[(-b)_+,(-b)_++\sqrt{t\epsilon(1-t)}]$, and therefore
\[
\frac{(q-t(1+\epsilon))_+}{1-t}\le h(t)\le\frac{(q+\epsilon-t)_+}{1-t}+\sqrt{\frac{\epsilon t}{1-t}} .
\]
Integrate, using $\int_0^r\frac{r-t}{1-t}dt=g(r)$ and $\int_0^1\sqrt{t/(1-t)}\,dt=\pi/2$. When $q+\epsilon>1$ the upper bound exceeds $1\ge\AM\ge D$ and holds trivially. The bounds on $g$ follow from its series.
\end{proof}

\paragraph{Asymptotics (\cref{thm:B}(ii)).}
We have $J(\nu_q)=\ln D(\nu_q)+(1-q)L$, since $\GM(\nu_q)=\epsilon^{1-q}$.

\emph{Lower bound.} Take $q=2/L$. Then $D\ge q^2/(2(1+\epsilon))$, so $J\ge L-2\ln L+\ln2-2+o(1)$.

\emph{Upper bound.} We split into three cases.
\begin{itemize}[leftmargin=*, itemsep=0pt, topsep=1pt]
\item[(A)] $q\ge\frac13$. Then $D\le1$, so $J\le\frac23L$.
\item[(B)] $q\le\epsilon^{1/8}$. For small $\epsilon$, $q+\epsilon\le2\epsilon^{1/8}\le\frac12$, so $D\le5\epsilon^{1/4}$ and $J\le\frac34L+\ln5$.
\item[(C)] $\epsilon^{1/8}\le q\le\frac13$. Here $\ln D\le2\ln q-\ln2-\ln(1-q)+O(\epsilon^{1/4})$ and $-\ln(1-q)\le1.22q$ on $[0,\frac13]$. Hence
\[
J\le L-\ln2+\max_q[2\ln q-q(L-1.22)]+o(1)=L-2\ln L+\ln2-2+o(1).
\]
\end{itemize}
Cases (A) and (B) are smaller by $\Omega(L)$. Hence $J^*=L-2\ln L+\ln2-2+o(1)$; dividing by $2\ln2$ gives (ii). Moreover $q^*L/2\to1$ for every maximizer. In case (C) the bound $f(q)=2\ln q-q(L-1.22)+L-\ln2+o(1)$ is strictly concave with maximizer $q_0=2/(L-1.22)$, and its maximum equals the lower bound $J(\nu_{2/L})$ up to $o(1)$. For $q=(1+x)q_0$, $f(q)-f(q_0)=-[2x-2\ln(1+x)]+o(1)\le-c_\delta+o(1)$ whenever $|x|\ge\delta$, with $c_\delta>0$. So for large $L$ no $q$ with $|x|\ge\delta$ can be a maximizer. We do not prove that the maximizing $q$ is unique, although it is unique numerically.

\paragraph{No rotation (\cref{thm:B}(iii)).}
Specht's theorem~\citep{specht1960theorie} says that $\sup\AM/\GM$ over $\cP_\epsilon$ is attained by a two-point measure on $\{\epsilon,1\}$. (The same KKT argument applies, since the first variation $y/\AM-\ln y$ is convex.) Parametrizing by $a=\AM$,
\[
\ln\frac{\AM}{\GM}=\max_{a\in[\epsilon,1]}\Big[\ln a+\frac{1-a}{1-\epsilon}L\Big],
\]
which is attained at $a=(1-\epsilon)/L$ with value $L-\ln L-1+o(1)$. Subtracting (ii) gives the saving.

\paragraph{Proof of \cref{cor:finite-n}.}
By \cref{thm:A}(iii) and its uniform probability bound, $\Delta_n-\Delta(\nu_n)\to0$ almost surely, uniformly over spectra with $\kappa\le1/\epsilon$, and $\Delta(\nu_n)\le\Delta^*(\epsilon)$. For quantiles of $\nu_{q^*}$, \cref{lem:A-continuity} gives equality in the limit.

\subsection{Normalization (N2): proof of \Cref{thm:Bprime}}
\label{app:Bprime}

By scale invariance, (N2) is equivalent to optimizing over $\cP_K=\{\nu:\supp\nu\subset[1,\infty),\ \AM(\nu)\le K\}$ with $K=1/\epsilon$. Let $\cP_{K,R}=\cP_K\cap\{\supp\nu\subset[1,R]\}$.

\begin{lemma}[Unimodality]
\label{lem:unimodal}
For a compactly supported, non-degenerate $\nu$ on $(0,\infty)$, the function
\[
u\mapsto e^{-u}(H*\ell)'(u)=\int H'(v)\,\frac{e^{-v}}{(1+e^{u-v})^2}\,dv
\]
is strictly decreasing, where $H(v)=\bar h_\nu(e^v)$.
\end{lemma}
\begin{proof}
Integration by parts gives $(H*\ell)'=\int H'(v)\ell(u-v)\,dv$, the boundary terms vanishing because $H$ is bounded. Moreover $e^{-u}\ell(u-v)=e^{-v}(1+e^{u-v})^{-2}$. Here $H'(v)e^{-v}=\bar h'(e^v)\ge0$ is bounded and not identically $0$, and the integral converges at both ends. For each $v$, $(1+e^{u-v})^{-2}$ is strictly decreasing in $u$.
\end{proof}

For $\beta>0$ let $\psi_\beta(e^u)=\Phi_\nu(e^u)/D-u-\beta e^u$. Then $\frac{d}{du}\psi_\beta(e^u)=g(u)/D$ with $g=H*\ell-D-D\beta e^u$, and $g'(u)=e^u[e^{-u}(H*\ell)'(u)-D\beta]$. By \cref{lem:unimodal}, $g'$ changes sign at most once, from $+$ to $-$, so $g$ has at most two zeros. Hence $\psi_\beta\circ\exp$ has at most two maximizers on any interval $[0,\ln R]$. If it has two, one of them is $u=0$, i.e.\ $y=1$.

\begin{lemma}[KKT with a mean constraint]
\label{lem:B-kkt}
If $\nu^*$ maximizes $J$ on $\cP_{K,R}$, there exist $\beta\ge0$ and $c$ such that $\Phi_{\nu^*}(y)/D(\nu^*)-\ln y-\beta y\le c$ on $[1,R]$, with equality on $\supp\nu^*$; moreover $\beta=0$ if $\AM(\nu^*)<K$.
\end{lemma}
\begin{proof}
By \cref{lem:B-gateaux} (after rescaling to $[1/R,1]$), $\nu^*$ maximizes the linear functional $\mu\mapsto\int\psi_0\,d\mu$ over $\cP_{K,R}$. Let $V(a)$ be the upper boundary of the closed convex hull of the curve $\{(y,\psi_0(y)):y\in[1,R]\}$. $V$ is concave, and $b^*=\int\psi_0\,d\nu^*=\max_{a\le K}V(a)$. Take $\beta=0$ if $\AM(\nu^*)$ is a global maximizer of $V$; otherwise take $\beta$ to be a supergradient of $V$ at $a^*=K$. Then $\psi_0(y)\le b^*+\beta(y-a^*)$ everywhere, with equality $\nu^*$-a.e., and hence on the support by continuity.
\end{proof}

\paragraph{Extremal structure.}
On $\cP_{K,R}$ a maximizer exists, and by the two displays above it is a two-point measure on $\{1,y_R\}$. In the scale-invariant form of \cref{thm:Bprime} the atom at $1$ is simply $\lambda_{\min}$; it lies at the bound $\epsilon\,\AM$ only if the mean constraint is active, which holds numerically but is not proved. For $\nu\in\cP_K$, moving the mass beyond $R$ to $R$ gives $\nu^{(R)}\in\cP_{K,R}$ with $J(\nu^{(R)})\to J(\nu)$, by monotone convergence for $\int\ln x$ and dominated convergence for $D$ (since $h\le\AM\le K$). Hence $\sup_{\cP_K}J$ is a limit of two-point values.

\paragraph{A refined two-point bound.}
For $q\le\frac13$, $\epsilon'\le\frac1{16}$ and $\nu=(1-q)\delta_{\epsilon'}+q\delta_1$,
\[
D(\nu)\le(1+\epsilon')g(q+\epsilon')+\epsilon'\Big[1+(q+\epsilon')\ln\tfrac1{\epsilon'}+4\sqrt{q+2\sqrt{\epsilon'}}\Big].
\]
This follows as in \cref{lem:B-twopoint}, now using $\sqrt{b^2+c}\le|b|+\min(\sqrt c,c/(2|b|))$ and splitting the integral at $|t-a|\le\sqrt{\epsilon'}$.

\paragraph{Asymptotics.}
Write $\nu_{q,y}=(1-q)\delta_1+q\delta_y$, so $J=\ln D-q\ln y$.

\emph{Lower bound.} With $q=1/L$ and $y=(K-1+q)/q$, \cref{lem:B-twopoint} gives $J\ge L-\ln L-1-\ln2-O(\ln L/L)$.

\emph{Upper bound.} Since $D\le\AM$, always $J\le\ln\frac{\AM}{\GM}=\ln(1-q+qy)-q\ln y$. Split into four cases.
\begin{itemize}[leftmargin=*, itemsep=0pt, topsep=1pt]
\item[(A)] $q\ge\frac13$. If $y\le K^{3/4}$, then $J\le\ln(1+y)\le\frac34L+1$; otherwise $J\le L-\frac13\ln y\le\frac34L$.
\item[(B)] $q<\frac13$ and $y<16$. Then $J\le\ln\AM\le\ln16$.
\item[(C)] $q<\frac13$ and $16\le y\le K/L^3$. Then $J\le\ln\AM\le\ln(1+y/3)\le L-3\ln L+O(1)$.
\end{itemize}
In (A)--(C), $J$ is smaller than $L-\ln L$ by a diverging amount.
\begin{itemize}[leftmargin=*, itemsep=0pt, topsep=1pt]
\item[(D)] $q<\frac13$ and $y>\max(16,K/L^3)$. The refined bound gives $D\le\frac{qK}{2(1-q)}(1+O(L^{-2}))$ when $q\ge L^3/K$, and with $q\ln y\ge q(L-3\ln L)$ this yields $J\le L-\ln L-1-\ln2+o(1)$. When $q<L^3/K$, $J=O(\ln L)$.
\end{itemize}
Dividing by $2\ln2$ gives the claim. For the unrotated case, $\GM\ge1$ and $\AM\le K$ give $\ln\frac{\AM}{\GM}\le\ln K$, approached by $q\to0$ with the AM constraint tight. \hfill$\square$

\subsection{Numerical verification}
\label{app:B-numerics}

\begin{itemize}[leftmargin=*, itemsep=1pt, topsep=1pt]
\item \emph{(N1) structure.} For $\epsilon\in\{10^{-2},\dots,10^{-12}\}$ and $m\in\{2,\dots,8\}$ atoms, $2304$ multistart optimizations never exceeded the best two-point value (by more than $7\cdot10^{-15}$ bit), and all collapsed to $\{\epsilon,1\}$.
\item \emph{First variation and convexity.} The formula of \cref{lem:B-gateaux} matched finite differences to relative error $6\cdot10^{-7}$ on $40$ random measures. Convexity in $\ln y$ held on $300$ random measures. At the optimal two-point spectrum, $\psi(\epsilon)=\psi(1)$ to $10^{-7}$, and $\psi$ is strictly smaller in the interior.
\item \emph{(N2).} Over $5$ values of $\epsilon$, $m\in\{2,3,4,6\}$ atoms and $48$ starts each, the optimizer never exceeded the best two-point value, which has one atom at the lower bound and a tight mean constraint.
\end{itemize}

\Cref{tab:N2} lists the (N2) values; \cref{tab:worst} lists the (N1) values.

\begin{table}[h]
\caption{Worst-case gaps under (N2), in bits. ``Asympt.''\ is \cref{thm:Bprime}; $\Delta^*_{N1}$ is shown for comparison.}
\label{tab:N2}
\centering
\small
\begin{tabular}{lcccccc}
\toprule
$\epsilon$ & $\Delta^*_{N2}$ & $q^*$ & $q^*L$ & asympt. & no rotation $\frac12\log_2\frac1\epsilon$ & $\Delta^*_{N1}$\\
\midrule
$10^{-2}$ & 0.950 & 0.179 & 0.82 & 0.999 & 3.322 & 0.463\\
$10^{-4}$ & 3.680 & 0.098 & 0.90 & 3.821 & 6.644 & 2.564\\
$10^{-6}$ & 6.734 & 0.066 & 0.91 & 6.850 & 9.966 & 5.272\\
$10^{-12}$ & 16.240 & 0.034 & 0.93 & 16.316 & 19.932 & 14.219\\
$10^{-48}$ & 75.083 & 0.009 & 0.97 & 75.111 & 79.726 & 72.000\\
\bottomrule
\end{tabular}
\end{table}

%% file: appendix/proofC.tex
\section{Proofs for \Cref{sec:typical}}
\label{app:C}

\paragraph{Scaling.}
We have $h_{c\nu}=c\,h_\nu$, $D(c\nu)=cD(\nu)$ and $\Delta(c\nu)=\Delta(\nu)$, because $T_{c\nu}(s)=T_\nu(s/c)$. Hence \cref{thm:A}, stated for $[\epsilon,1]$, applies to any compactly supported spectrum after rescaling.

\paragraph{Rank-$r$ interlacing.}
For Hermitian $A$ and Hermitian $E$ of rank $\le r$, with eigenvalues in increasing order,
\[
\gamma_{i-r}(A)\le\gamma_i(A+E)\le\gamma_{i+r}(A),
\]
where out-of-range indices are read as $\mp\infty$. Indeed, on $U\cap\ker E$, with $U$ the span of the first $i+r$ eigenvectors of $A$ (a subspace of dimension $\ge i$), we have $x^\top(A+E)x\le\gamma_{i+r}(A)\|x\|^2$; apply Courant--Fischer, and repeat with $A+E$ and $-E$. Consequently, if $A$ and $A+E$ are positive definite with all eigenvalues $\ge\epsilon$,
\begin{equation}
\big|\tr A^{-1}-\tr(A+E)^{-1}\big|\le r/\epsilon .
\label{eq:trinv}
\end{equation}

\subsection{Proof of \Cref{prop:outliers}}

\emph{(i)} For each $n$, build diagonal $\Lambda_n,\Lambda_n'$ that share $\lfloor(1-\eta)n\rfloor$ entries at the quantiles of $\rho$, with the remaining entries at the quantiles of $\alpha$ (resp.\ $\beta$). Then $\nu_n\Rightarrow\nu$, $\nu_n'\Rightarrow\nu'$, and the two matrices differ in $r\le\eta n+1$ entries. Use the same Haar $V_n$ for both. The leading $k\times k$ blocks differ by a matrix of rank $\le r$, and all their eigenvalues are $\ge\epsilon$, so by \eqref{eq:trinv} $|m_k-m_k'|\le r/(\epsilon k)$, deterministically. For $k=\lceil tn\rceil$, Step~2 of \cref{app:A} (after rescaling) gives $m_k\to1/h_\nu(t)$ and $m_k'\to1/h_{\nu'}(t)$ almost surely. Hence $|1/h_\nu(t)-1/h_{\nu'}(t)|\le\eta/(\epsilon t)$. The case $t=1$ is direct: $|\int x^{-1}d(\nu-\nu')|\le\eta/\epsilon$.

\emph{(ii)} $T_\rho(s)\le b/(s+b)$ and $T_\beta\le1$, so $t=T_{\nu'}(s_t)\le(1-\eta)\frac{b}{s_t+b}+\eta$. This gives $s_t\le\frac{b(1-t)}{t-\eta}$ and $h_{\nu'}(t)\le\frac{bt}{t-\eta}$. The same holds for $\nu$.

\emph{(iii)} On $[2\eta,1]$, (ii) gives $h_\nu,h_{\nu'}\le2b$, so $|h_\nu-h_{\nu'}|=h_\nu h_{\nu'}|1/h_\nu-1/h_{\nu'}|\le4b^2\eta/(\epsilon t)$, which integrates to $\frac{4b^2\eta}{\epsilon}\ln\frac1{2\eta}$. On $[0,2\eta]$, $h\le\AM$, so the integral is at most $2\eta(\AM(\nu)+\AM(\nu'))$.

\emph{(iv)} Here $\nu=\rho$, $\nu'=(1-\eta)\rho+\eta\delta_\mu$, and we write $h'=h_{\nu'}$.

\emph{Lower bound.} For $t<\eta$, $T_{\nu'}(s)\ge\eta\mu/(s+\mu)$ gives $s_t\ge\mu(\eta-t)/t$, hence $h'(t)\ge\mu(\eta-t)$. Integrating over $(0,\eta)$, and using $D(\nu)\le\AM(\rho)\le b$,
\[
D(\nu')-D(\nu)\ge\tfrac12\eta^2\mu-b .
\]

\emph{Upper bound for $t<\eta$.} At $s=s_t$,
\[
t=(1-\eta)T_\rho(s)+\frac{\eta\mu}{s+\mu}\le\frac bs+\frac{\eta\mu}{s+\mu}.
\]
Multiplying $t-b/s\le\eta\mu/(s+\mu)$ by $s+\mu$ gives $ts\le\mu(\eta-t)+b+b\mu/s$. The lower bound on $s_t$ gives $\mu/s_t\le t/(\eta-t)$, so
\[
h'(t)\le\frac{\mu(\eta-t)+b+bt/(\eta-t)}{1-t}.
\]

\emph{Integrating.} Fix $\tau=\eta\mu^{-1/2}\le\eta$ and split $(0,1)$ into three ranges.
\begin{itemize}[leftmargin=*, itemsep=0pt, topsep=1pt]
\item On $(0,\eta-\tau)$, the display gives $\int h'\le\frac{1}{1-\eta}\big[\frac12\mu\eta^2+b\eta+\frac12b\eta\ln\mu\big]$, using $\int_0^{\eta-\tau}\frac{t\,dt}{\eta-t}\le\eta\ln\frac\eta\tau$.
\item On $[\eta-\tau,2\eta]$, $h'$ is non-increasing, so $h'\le h'(\eta-\tau)\le\frac{\mu\tau+b+b\eta/\tau}{1-\eta}$. The interval has length $\le2\eta$, so the contribution is at most $\frac{2}{1-\eta}\big[\eta^2\mu^{1/2}+b\eta+b\eta\mu^{1/2}\big]$.
\item On $[2\eta,1]$, (iii) gives $\int h'\le D(\nu)+\frac{4b^2\eta}{\epsilon}\ln\frac1{2\eta}$.
\end{itemize}
Adding the three ranges,
\[
D(\nu')-D(\nu)\le\frac{\frac12\eta^2\mu}{1-\eta}+O\big(\eta^2\mu^{1/2}+b\eta\mu^{1/2}+b\eta\ln\mu\big)+\frac{4b^2\eta}{\epsilon}\ln\frac1{2\eta}.
\]
As $\eta\to0$ and $\eta^2\mu\to\infty$, every error term is $o(\eta^2\mu)$:
\begin{itemize}[leftmargin=*, itemsep=0pt, topsep=1pt]
\item $\eta^2\mu^{1/2}/(\eta^2\mu)=\mu^{-1/2}\to0$;
\item $b\eta\mu^{1/2}/(\eta^2\mu)=b/(\eta^2\mu)^{1/2}\to0$;
\item $b\eta\ln\mu/(\eta^2\mu)=b\ln\mu/(\eta\mu)\to0$;
\item the last term is $O(\eta\ln\frac1\eta)=o(1)$.
\end{itemize}
The lower bound is also $\frac12\eta^2\mu(1-o(1))$. Hence $D(\nu')-D(\nu)=\frac12\eta^2\mu\,(1+o(1))$.

\emph{Interpretation.} The factor $\frac12$ comes from $h'(t)\approx\mu(\eta-t)$ on $t<\eta$: the first $\approx\eta n$ pivots absorb the outlier subspace. When instead $\eta^2\mu\to0$, (iii) shows that the outliers have no effect in the limit. The synthetic numbers in \cref{sec:typical} sit between these regimes and are close to the $\frac12$ formula.

\subsection{Proof of \Cref{prop:spikes}}

Let $\Sigma=V^\top\bar\Lambda V$ and $\Sigma^0=V^\top\Lambda^0V$, where $\Lambda^0$ replaces the spikes, indexed by $S_n$, by $1$. Write $\bar M,M^0$ for the leading $k\times k$ blocks, $\bar m_k,m_k^0$ for their normalized inverse traces, and $\pi_k$ for the pivots of $\Sigma$. We use the Gaussian case of \cref{lem:A-hw} in the form
\begin{equation}
\Prob(|\chi^2_d-d|>s)\le2\exp[-c\min(s^2/d,\,s)].
\label{eq:chi2}
\end{equation}

\emph{First claim ($k\ge\delta n$).} \Cref{lem:A-sphere} holds for any $\Lambda$, and $1/\bar\gamma_i\in(0,1/\epsilon]$. Step~1 of \cref{app:A} only uses $\|\bar{\mathcal A}-\bar m_kI\|\le1/\epsilon$, so $\max_{k\ge\delta n}|(\bar M^{-1})_{kk}-\bar m_k|\to0$ almost surely. Since $\bar M-M^0\succeq0$ has rank $\le r_n$, \eqref{eq:trinv} gives $|\bar m_k-m_k^0|\le r_n/(\epsilon\delta n)\to0$. Steps~2--3 of \cref{app:A} for $\Lambda^0$ give $m_k^0\to1/h_{\nu^0_n}(k/n)$ uniformly. Since $1/h\ge1$, eventually $(\bar M^{-1})_{kk}\ge\frac12$, and $|\pi_k-h|\le2|(\bar M^{-1})_{kk}-1/h|\to0$.

\emph{Second claim, middle range.} Let $K_n=\lceil C_1(r_n+\ln n)\rceil$ with $C_1=\max(8,32/c)$. Since $M^0\preceq I$ and $\bar M=M^0+P$ with $P\succeq0$ of rank $\le r_n$, we have $\bar\gamma_i\le1$ for $i\le k-r_n$. Hence, by \cref{lem:A-sphere},
\[
(\bar M^{-1})_{kk}\overset{d}{=}\sum_i\frac{w_i^2}{\bar\gamma_i}\ge1-\sum_{i>k-r_n}w_i^2 .
\]
The index set is determined by $\bar\gamma$ and is therefore independent of $w$, so $\sum_{i>k-r_n}w_i^2\overset{d}{=}\chi^2_{r_n}/(\chi^2_{r_n}+\chi'^2_{k-r_n})$. For $k\ge8r_n$, \eqref{eq:chi2} gives
\[
\Prob(\chi^2_{r_n}>\chi'^2_{k-r_n})\le\Prob(\chi^2_{r_n}\ge k/4)+\Prob(\chi'^2_{k-r_n}\le k/4)\le4e^{-ck/8}.
\]
A union bound over $k\in[K_n,n]$ costs $4ne^{-cK_n/8}\le4n^{-3}$. Hence almost surely $\pi_k\le2$ for all $k\ge K_n$ eventually.

\emph{Second claim, early range.} We have $\pi_k\le\Sigma_{kk}$, so
\[
\sum_{k<K_n}\pi_k\le\sum_j\bar\lambda_j\|V_{j,<K_n}\|^2\le(K_n-1)+\sum_{j\in S_n}\mu_j\|V_{j,<K_n}\|^2 .
\]
Each row of $V$ is uniform on the sphere, so $\|V_{j,<K}\|^2\overset{d}{=}\chi^2_{K-1}/\chi^2_n$. By \eqref{eq:chi2} and a union bound over $j$, almost surely $\max_{j\in S_n}\|V_{j,<K_n}\|^2\le\theta_n=O((r_n+\ln n)/n)$ eventually. Hence
\[
\frac1n\sum_{k<K_n}\pi_k\le\frac{K_n}{n}+\frac{\theta_n}{n}\sum_j\mu_j\to0 .
\]

\emph{Combining the ranges.} The middle range contributes at most $2\delta$, the part of $D(\nu^0_n)$ with $t<\delta$ at most $\delta$, and the Riemann error at most $2/n$. So $\limsup|\frac1n\sum\pi_k-D(\nu_n^0)|\le3\delta$ for every $\delta$. Finally, $\GM(\bar\Lambda)=\GM(\nu^0_n)\exp(\frac1n\sum\ln\mu_j)$ and $D\ge\epsilon$ give the formula for $\Delta$.

\emph{Third claim.} Let $r_n=o(n)$ and $\sum_j\mu_j\ge cn^2$ for some $c\in(0,1]$. Then $\pi_1=\Sigma_{11}\ge X:=\sum_{j\in S_n}\mu_jV_{j1}^2$, and the first column of $V$ is uniform on the sphere. Since $\E V_{j1}^4=\frac{3}{n(n+2)}$ and $\E V_{i1}^2V_{j1}^2=\frac{1}{n(n+2)}$ for $i\ne j$, we have $\E X^2\le3(\E X)^2$ with $\E X=\frac1n\sum_j\mu_j$, so the Paley--Zygmund inequality gives $\Prob(X\ge\frac12\E X)\ge\frac1{12}$. On this event $\pi_1/n\ge c/2$. By the first claim and $h\le1$, almost surely $\frac1n\sum_{k\ge\delta n}\pi_k\ge D(\nu^0_n)-\delta-o(1)$. With $\delta=c/4$, the event $\frac1n\sum_k\pi_k\ge D(\nu^0_n)+\frac c4-o(1)$ therefore has probability at least $\frac1{12}-o(1)$, and on it the formula for $\Delta$ fails by at least $\frac12\log_2(1+\frac c4)-o(1)$, because $D\le1$. For a single spike, $\pi_1\approx\mu/n$.

\subsection{Proof of \Cref{prop:logscale}}

Let $L=\ln\kappa$. Substituting $t=T(s)$ gives the representation
\begin{equation}
D(\nu)=\int_0^\infty\bar h_\nu(s)B_\nu(s)\,ds,\qquad \bar h_\nu=\frac{T_\nu}{A_\nu}=\frac{sT_\nu}{1-T_\nu}.
\label{eq:Dints}
\end{equation}
Substituting $w=Lu$, as $L\to\infty$ and for each fixed $s>0$:
\begin{itemize}[leftmargin=*, itemsep=0pt, topsep=1pt]
\item $LT_L(s)=\int_0^Lf(w/L)\frac{dw}{1+se^w}\to f(0)\ln(1+\frac1s)$;
\item $LB_L(s)\to\frac{f(0)}{s(1+s)}$;
\item $A_L(s)\to\frac1s$.
\end{itemize}
(Each by dominated convergence.) The integrand of $L^2D$ therefore converges to $f(0)^2\frac{\ln(1+1/s)}{1+s}$, and $\int_0^\infty\frac{\ln(1+1/s)}{1+s}ds=\int_0^1\frac{-\ln(1-y)}{y}dy=\zeta(2)$.

To exchange limit and integral, split $(0,\infty)$ at a fixed $s_0$ and at $e^{-La}$ with $a=1/(8\|f\|_\infty)$:
\begin{itemize}[leftmargin=*, itemsep=0pt, topsep=1pt]
\item On $[s_0,\infty)$, $A_L\ge\frac1{1+s}$ gives the integrable majorant $\|f\|_\infty^2/s^2$.
\item On $[e^{-La},s_0]$, $T_L\le\frac12$, so the integrand is at most $2\|f\|_\infty^2\ln(1+1/s)$, whose integral over $(0,s_0]$ vanishes as $s_0\to0$.
\item On $(0,e^{-La}]$, the contribution is at most $L^2\bar h_L(e^{-La})\le2L^2e^{-La}\to0$, since $h$ is decreasing.
\end{itemize}
Hence $L^2D\to f(0)^2\zeta(2)$. Moreover $L\,\AM=\int_0^Lf(w/L)e^{-w}dw\to f(0)$ and $\ln\GM=-L\,\E[U]$, which give the two formulas.

\paragraph{The log-uniform spectrum.}
Here $T(s)=\frac1L\ln\frac{1+s}{s+e^{-L}}$, so $s_t=\frac{1-e^{-(1-t)L}}{e^{tL}-1}$ exactly, and
\[
L^2D=\int_0^L\frac{u\,(1-e^{-(L-u)})}{(1-u/L)(e^u-1)}\,du=\zeta(2)+\frac{2\zeta(3)}{L}+\frac{6\zeta(4)}{L^2}+O(L^{-3}).
\]
For $u\le L/2$, expand $(1-u/L)^{-1}$ to second order with remainder $\le2(u/L)^3$ and use $\int_0^\infty\frac{u^{m+1}}{e^u-1}du=(m+1)!\zeta(m+2)$; the range $u>L/2$ is exponentially small. Hence
\[
\Delta_\rot=\tfrac14\log_2\kappa-\log_2L+\tfrac12\log_2\zeta(2)+\frac{c_1}{L}+\frac{c_2}{L^2}+O(L^{-3}),
\]
with $c_1=\frac{\zeta(3)}{\zeta(2)\ln2}\approx1.0543$ and $c_2=\frac{\pi^2/5-(\zeta(3)/\zeta(2))^2}{\ln2}\approx2.0773$. The unrotated gap is exactly $\frac14\log_2\kappa-\frac12\log_2L+\frac12\log_2(1-e^{-L})$. \Cref{tab:logu} compares these expressions with direct numerical solution of $T(s)=t$.

\begin{table}[h]
\caption{Log-uniform spectrum: rotated gap in bits.}
\label{tab:logu}
\centering
\small
\begin{tabular}{lcccc}
\toprule
$\kappa$ & numerical & exact formula & expansion to $c_1/L$ & expansion to $c_2/L^2$\\
\midrule
$10^4$ & 0.634 & 0.6338 & 0.5922 & 0.6166\\
$10^6$ & 1.646 & 1.6457 & 1.6300 & 1.6409\\
$10^8$ & 2.865 & 2.8647 & 2.8568 & 2.8630\\
$10^{12}$ & 5.578 & 5.5779 & 5.5747 & 5.5775\\
\bottomrule
\end{tabular}
\end{table}

%% file: appendix/proofS.tex
\section{Proof of \Cref{prop:S}}
\label{app:S}

\paragraph{Rate decomposition.}
By the entropy part of \cref{ass:H} and $\prod_i\tilde\alpha_i=1$, the rate is
\[
R(\alpha)=\frac1n\sum_iH(Z_i)=\log_2\frac1\alpha+\hd_{\rm col}(W)+o(1).
\]
By the distortion part, $\log_2\frac1\alpha=\frac12\log_2\frac{\frac1{12n}\sum_i\tilde\alpha_i^2\pi_i}{\dist}+o(1)$. Combining the two,
\[
R=\tfrac12\log_2\frac{\frac1n\sum_i\tilde\alpha_i^2\pi_i}{12\dist}+\hd_{\rm col}(W)+o(1).
\]
For PlainWaterSIC, $\tilde\alpha_i^2\pi_i=(\det\Sigma)^{1/n}$ for every $i$, so the first term is basis independent. For GPTQ, $\tilde\alpha_i=1$, so the first term involves $\AM(\pi)$. Subtracting the rates before and after the rotation $(W,\Sigma)\mapsto(WQ,Q^\top\Sigma Q)$ gives both limits.

\paragraph{Identity and bound.}
By the definition of negentropy, $\hd(W_i)=\frac12\log_2(2\pi e\sigma_i^2)-\Neg(W_i)$. Hence
\[
\hd_{\rm col}(W)=\tfrac12\log_2\big(2\pi e\,\AM(\sigma^2)\big)-\WS(W).
\]
The column variances of $WQ$ are $q_i^\top Cq_i$, where $C$ is the row covariance, so their sum $\tr C$ is rotation invariant and $\AM(\sigma^2)$ is unchanged. Subtracting gives $\Cw(Q)=\WS(W)-\WS(WQ)$. Since $\WS\ge0$, with equality iff all columns are Gaussian with equal variances, the bound follows. \hfill$\square$

\paragraph{When does \cref{ass:H} hold?}
\citet[\S3.4]{lifar2026thesis} prove both parts for i.i.d.\ Gaussian rows. The distortion part uses the flatness factor of the lattice. The entropy part uses three ingredients: SIC decisions of the form $Z_i=\mathrm{round}(W_i/(\alpha\tilde\alpha_i)+e_i)$ with $e_i$ bounded and \emph{independent of $W_i$}; R\'enyi's high-resolution entropy theorem; and continuity of differential entropy under small independent perturbations. For correlated columns, $e_i$ depends on $W_{i+1:n}$, which is correlated with $W_i$. We expect the conclusion to persist whenever the conditional density of $W_i$ given $W_{i+1:n}$ is regular, since $W_i+\alpha e_i$ is then a mixture of slightly shifted conditional densities. We leave a proof to future work.

\paragraph{Random rotations.}
For a Haar rotation, each column of $WQ$ is a random one-dimensional projection of the rows of $W$. Under suitable conditions most such projections are close to Gaussian~\citep{diaconis1984asymptotics}, and their variances concentrate at $\AM(\sigma^2)$. We therefore expect $\WS(WQ)\to0$ and $\Cw(Q)\to\WS(W)$, but we have not verified the conditions for real weights. Measured after one round of randomized Hadamard (\cref{app:e4}), the variance term of $\WS(WQ)$ drops below $0.001$ bit, while the negentropy term remains $0.02$--$0.04$ bit on q/k/v inputs. The ratio $\Cw/\WS(W)$ has median $0.31$--$0.88$ depending on the layer type.

%% file: appendix/experiments.tex
\section{Experimental Details and Additional Results}
\label{app:exp}

\subsection{Setup}

\paragraph{Models and data.}
We use Qwen3-1.7B, Qwen3-8B-Base and Llama-3.1-8B-Instruct. The E4 and E6 runs pin the Hugging Face revisions \texttt{70d244c} (Qwen3-1.7B) and \texttt{49e3418} (Qwen3-8B-Base). For Llama they use a public mirror whose weight, tokenizer and config files are byte-identical to the official release. The E3 Hessians were computed from locally stored copies of the same models; their revisions were not recorded, but the layers recomputed in E6 reproduce the earlier values to all printed digits. Calibration uses $128$ sequences of $2048$ tokens from the WikiText-2 training split~\citep{merity2017pointer} (seed $0$). Perplexity is measured on the test split.

\paragraph{Hessians.}
$\Sigma=\E[xx^\top]$ is accumulated in float64. The damping is $\Sigma+\lambda I$ with $\lambda=d\cdot\overline{\diag\Sigma}$ and $d\in\{0,10^{-3},10^{-2}\}$. With $d=0$, layers with $\lambda_{\min}\le10^{-12}\lambda_{\max}$ are numerically singular and are skipped ($18$ of $96$).

\paragraph{Rotations.}
Haar rotations are sampled by QR with sign correction. Randomized Hadamard rotations are $DH_n$ (one round) and $D_1H_nD_2H_n$ (two rounds), where $D$ is a random sign diagonal and $H_n$ is the normalized $\mathrm{Sylvester}(2^k)\otimes\mathrm{Paley}(m)$ matrix with $m\in\{1,12,28\}$. This covers $n=6144,12288,14336$, with orthogonality error $\le2\cdot10^{-13}$. Gaps are computed from the pivots in the GPTQ order (\cref{app:A}).

\paragraph{Numerical checks.}
For every basis, $\frac1n\sum_k\ln\pi_k=\frac1n\ln\det\Sigma$ to $3\cdot10^{-10}$. The quadrature for $D$ integrates to $1\pm10^{-15}$. The annealed approximation of \citet{ordentlich2026hrqmm2} differs from $D(\nu_{\rm emp})$ by $\le0.008$ bit (mostly $<10^{-3}$).

\paragraph{Compute.}
The Hessian experiments (E3) took $5$--$16$ minutes per model on one 80GB GPU. The rate--distortion, perplexity and lattice-reduction experiments (E4, E6) ran as one job queue on two 80GB GPUs for $3$ hours. All synthetic experiments ran on CPU.

\subsection{Synthetic spectra (E1)}

\begin{table}[h]
\caption{Error of the limit of \cref{thm:A}, $\Delta_n-\Delta(\nu_n)$ in bits, averaged over seeds. ``had1'' is one round of randomized Hadamard applied to a \emph{diagonal} $\Sigma$, ``had2'' two rounds.}
\label{tab:e1}
\centering
\small
\begin{tabular}{lcccccccc}
\toprule
spectrum & $\Delta(\nu)$ & no rotation & \multicolumn{2}{c}{Haar} & \multicolumn{2}{c}{had2} & \multicolumn{2}{c}{had1}\\
 & & & $n=256$ & $8192$ & $256$ & $8192$ & $256$ & $8192$\\
\midrule
two-point $q=0.10$, $\epsilon=10^{-3}$ & 0.890 & 2.830 & +0.0167 & $-$0.0001 & +0.0305 & $-$0.0000 & $-$0.0221 & $-$0.0751\\
two-point $q=0.22$, $\epsilon=10^{-4}$ & 2.563 & 4.090 & +0.0008 & $-$0.0004 & +0.0044 & $-$0.0004 & $-$0.0385 & $-$0.1469\\
power law $\beta=3$, $\epsilon=10^{-4}$ & 0.397 & 1.043 & +0.0040 & +0.0002 & $-$0.0023 & +0.0002 & $-$0.0072 & $-$0.0239\\
log-uniform, $\epsilon=10^{-4}$ & 0.634 & 1.720 & +0.0093 & +0.0003 & +0.0082 & +0.0003 & $-$0.0186 & $-$0.0394\\
Marchenko--Pastur, ratio $0.25$ & 0.0025 & 0.099 & +0.0008 & +0.0000 & +0.0008 & +0.0000 & $-$0.0000 & $-$0.0000\\
\bottomrule
\end{tabular}
\end{table}

Under Haar rotations the relative error of the mean pivot decays with fitted exponents $-0.81$ to $-1.14$, and the pointwise relative r.m.s.\ error with exponents $-0.51$ to $-0.59$. Two rounds of Hadamard are indistinguishable from Haar for $n\ge2048$. At $n=256$ the seed-to-seed standard deviation of both reaches $0.03$ bit, which covers their difference in \cref{tab:e1}; for the two-point spectra the order is reversed at $n=512$. One round applied to a diagonal $\Sigma$ does \emph{not} follow \cref{thm:A}: the random signs commute with $\Sigma$, so only a fixed Hadamard and a permutation remain. The resulting gaps are $0.02$--$0.15$ bit \emph{smaller} than the Haar limit and do not converge to it. This setting needs separate treatment; on real (non-diagonal) Hessians the effect is small (\cref{sec:exp}).

\subsection{Real Hessians (E3)}
\label{app:e3}

\begin{table}[h]
\caption{Gaps in bits at damping $10^{-2}$; each cell is the median over $8$ blocks. ``natural'' is unrotated GPTQ in the natural order, ``PCA'' the worst basis, and $\Delta^*(\kappa)$ the worst case at the layer's condition number.}
\label{tab:e3}
\centering
\small
\begin{tabular}{llcccccccc}
\toprule
model & input & $\kappa$ & $\Delta(\nu_{\rm emp})$ & Haar & Hadamard & natural & act-order & PCA & $\Delta^*(\kappa)$\\
\midrule
Qwen3-1.7B & q/k/v & 5.1e4 & 0.039 & 0.041 & 0.039 & 0.238 & 0.076 & 1.15 & 3.48\\
 & o\_proj & 2.9e4 & 0.081 & 0.082 & 0.081 & 0.271 & 0.192 & 1.07 & 3.17\\
 & gate/up & 5.7e4 & 0.026 & 0.026 & 0.026 & 0.130 & 0.023 & 0.85 & 3.55\\
 & down\_proj & 2.7e4 & 0.057 & 0.057 & 0.056 & 0.341 & 0.229 & 0.78 & 3.11\\
Qwen3-8B-Base & q/k/v & 6.8e4 & 0.040 & 0.040 & 0.040 & 0.261 & 0.077 & 1.08 & 3.65\\
 & o\_proj & 3.7e4 & 0.070 & 0.071 & 0.075 & 0.212 & 0.146 & 1.00 & 3.28\\
 & gate/up & 7.0e4 & 0.020 & 0.021 & 0.021 & 0.072 & 0.013 & 0.65 & 3.67\\
 & down\_proj & 1.3e4 & 0.044 & 0.044 & 0.043 & 0.295 & 0.176 & 0.67 & 2.71\\
Llama-3.1-8B-I. & q/k/v & 4.9e4 & 0.030 & 0.030 & 0.030 & 0.180 & 0.085 & 0.85 & 3.46\\
 & o\_proj & 4.4e4 & 0.091 & 0.092 & 0.093 & 0.247 & 0.147 & 1.26 & 3.40\\
 & gate/up & 9.3e3 & 0.018 & 0.018 & 0.018 & 0.030 & 0.012 & 0.49 & 2.52\\
 & down\_proj & 1.2e4 & 0.024 & 0.024 & 0.024 & 0.137 & 0.065 & 0.57 & 2.69\\
\bottomrule
\end{tabular}
\end{table}

\paragraph{Without damping.}
At $d=0$, which is the setting of \citet{ordentlich2026hrqmm2}, the Haar gap of non-singular layers has median $0.070$, $0.054$ and $0.029$ bit for the three models. Between $71\%$ and $78\%$ of these layers are below $0.1$ bit. The layers above $0.1$ bit are mostly o\_proj inputs: $7$ of $8$ blocks for Llama, with a maximum of $0.63$ bit.

\paragraph{Two rounds of Hadamard.}
The ``Hadamard'' column of \cref{tab:e3} is one round. Over all $270$ (layer, damping) pairs, two rounds are within $0.01$ bit of the prediction on $98\%$, and every exception is an o\_proj input. On the other input types the two-round gap exceeds the prediction by $10^{-4}$ to $0.006$ bit, as under Haar. On the $72$ o\_proj pairs it lies between $0.048$ bit below and $0.0003$ bit above the prediction, and it is smaller than the Haar gap on every one of them. The cause is that for $n=2^k$ the Sylvester matrix factors as $H_{n/128}\otimes H_{128}$, aligned with the $128$-dimensional attention heads, as the next paragraph shows.

\paragraph{Random permutations (E8).}
We collected the same Hessians again; the Haar and two-round gaps agree with E3 exactly. On the $24$ o\_proj inputs we compared variants of two rounds, with $8$ seeds each, at damping $10^{-2}$; the other dampings give the same picture. Let $d$ be the gap minus the Haar gap. Two rounds give $d<0$ on all $24$ layers, with median $-0.0025$ bit.
\begin{itemize}[leftmargin=*, itemsep=0pt, topsep=1pt]
\item \emph{Breaking the alignment removes the advantage.} With a uniformly random permutation of the inputs before $D_1$, $d<0$ on only $11$ of $24$ layers, and the median is $0.0000$ bit. A uniformly random GPTQ order after the rotation has the same effect ($16$ of $24$, median $-0.0001$ bit).
\item \emph{Permutations that keep the blocks do not.} Permuting the coordinates within each head, or permuting whole KV groups, leaves the median $d$ unchanged ($102\%$ and $100\%$ of it). Permuting whole heads, which separates heads of the same group, leaves $58\%$.
\item \emph{Fake Haar rotations.} The rotation of \cref{prop:liberating}(ii) and its independent version give gaps above $\Delta(\nu_{\rm emp})$ on all $24$ layers at every damping, by at least $0.0003$ bit. Two rounds are below it on $21$ of the $24$ layers.
\end{itemize}
The fake Haar gaps are below the Haar gap on $21$ of $24$ layers, so our pre-registered sign test, which had no threshold on the size of the effect, classes them as different from Haar. The median difference is only $0.0001$ bit, against a Haar seed standard deviation of $0.0003$ bit. The q/k/v inputs, which have no head structure, show the same small difference. On all input types other than o\_proj, the fake Haar gaps are within $0.005$ bit of Haar on at least $98\%$ of the (layer, damping) pairs.

Last, we kept only the diagonal head blocks of each o\_proj Hessian. At the real $n$, these block-diagonal matrices reproduce the advantage, with median $d=-0.0025$ bit. We then drew blocks with replacement to grow $n$ from $1024$ to $16384$. The median $d$ fell from $-0.0029$ to $-0.0003$ bit, with fitted slope $-0.87$ in $\log n$. KV-group blocks give $-0.0030$ to $-0.0005$ bit, with slope $-0.67$. So the advantage of two rounds on o\_proj inputs appears to be a finite-$n$ effect of the head alignment, and gives no evidence that they violate condition (L) of \cref{app:A-liberating}.

\paragraph{Act-order versus rotation (E7).}
We collected the same Hessians again (they agree with E3 to $10^{-15}$ relative error) and computed their eigenvectors. At damping $10^{-2}$, act-order beats the Haar rotation on $26$ of the $96$ layers. We tested three explanations, with criteria fixed before the run.
\begin{itemize}[leftmargin=*, itemsep=0pt, topsep=1pt]
\item \emph{Axis-aligned outliers.} For the top eigenvector $v_1$, the ratio $\|v_1\|_\infty/\|v_1\|_2$, divided by its typical value $\sqrt{2\ln n/n}$ for a random unit vector, has median $4.9$ on these layers and $5.9$ on the others. Across (winning, losing) pairs of layers it is larger on the winning layer only $39\%$ of the time ($54\%$ within gate/up inputs). Measures based on the top $1$ to $16$ eigenvectors give $33$--$47\%$. Exact alignment would in fact favour rotation: a coordinate uncorrelated with all others has pivot $\Sigma_{jj}$ in every order.
\item \emph{Outlier eigenvalues.} Clipping the top $r$ eigenvalues to the $(r+1)$-th, with the eigenvectors kept, leaves all $26$ layers winning, both for $r$ the number of eigenvalues above $100$ times the median and for $r=\lceil0.01n\rceil$. The rank correlation of the advantage before and after clipping is at least $0.99$. With $r$ the number above $10$ times the median, $20$ of the $26$ still win.
\item \emph{Ordering.} The order matters on every layer: reversing act-order costs a median of $0.25$ bit where act-order wins and $0.31$ bit elsewhere. Ordering by leverage in the top eigenspace gives within $0.02$ bit of act-order on every layer. So the ordering effect does not single out these layers.
\end{itemize}
One quantity does separate the layers: the fraction of the spectral spread that is visible on the diagonal, that is, $\frac12\log_2(\AM/\GM)$ of $\diag\Sigma$ divided by the same quantity for the eigenvalues. Its median is $0.31$ on the winning layers and $0.49$ on the others, and it is smaller on the winning layer for $78\%$ of the pairs. Within each input type that has winning layers the figure is $75$--$84\%$, but one side then has only $3$ to $5$ layers. We had listed it only as an auxiliary measure, so this finding is exploratory. The results are similar at dampings $10^{-3}$ and $0$.

The direction may seem odd, since act-order sorts by the diagonal. A decomposition we made after seeing the data suggests why. Write the advantage as $\Delta_{\rm act}-\Delta_{\rm Haar}=(\Delta_{\rm nat}-\Delta_{\rm Haar})-(\Delta_{\rm nat}-\Delta_{\rm act})$, where ``nat'' is GPTQ in the native order. The first term, the excess of the native basis over a rotation, has median $0.057$ bit on the winning layers and $0.186$ bit on the others. It is smaller on the winning layer for $90\%$ of the pairs, and for $83$--$84\%$ within each input type. The second term, the gain from ordering, is not larger on the winning layers: the medians are $0.066$ and $0.104$ bit, and it is larger for only $29\%$ of the pairs (o\_proj inputs are the exception, at $82\%$). So act-order wins where the native basis is already nearly as good as a rotated one, not where ordering helps more. A rotation flattens the diagonal, since a Haar-rotated $\Sigma$ has all diagonal entries close to $\AM(\nu)$. Where the native diagonal is already flat, a rotation therefore has less to gain. Act-order uses only the order of the diagonal entries, not their spread. Accordingly, the diagonal fraction has rank correlation $0.75$ with the first term. This is a description, not a test, because the two terms add up to the advantage by construction.

\paragraph{Distortion side of the high-rate formula.}
With per-channel AbsMax INT-$b$ quantization, the ratio of the measured layer loss to $\frac{s^2}{12}\sum_k\pi_k$ after a rotation is $1.000$ (range $[0.996,1.004]$) for $b\in\{3,4,8\}$, and $0.81$--$0.85$ for $b=2$. Without rotation it already deviates at $b=3$.

\subsection{Rate side, weight-side cost and perplexity (E4)}
\label{app:e4}

\paragraph{Methods.}
\begin{itemize}[leftmargin=*, itemsep=0pt, topsep=1pt]
\item \emph{Entropy-coded GPTQ} (Huffman-GPTQ): common step, no clipping. The rate is the mean per-column empirical code entropy with Miller--Madow correction (ideal entropy coding, codebooks not counted).
\item \emph{PlainWaterSIC}: steps $\alpha_i=\alpha\,U_{ii}^{-1}\GM(U_{11},\dots,U_{nn})$.
\item \emph{Rate gap}: horizontal distance between rate--distortion curves measured on a grid of nominal rates $1$--$7$ bits, on $8$ blocks $\times$ $4$ input types per model.
\item \emph{Fixed rate}: INT-$b$ with group-128 AbsMax scales and clipping.
\item \emph{Perplexity}: sequential pipeline in which each block's Hessians are collected on the outputs of the already-quantized blocks, with rate matched to $\pm0.01$ bit by a secant search on the step.
\end{itemize}

\begin{table}[h]
\caption{Equal-distortion rate gap at $R=4$ (bits; median over $8$ blocks), predicted / measured. ``Weight-side cost'' is $\Cw$ for one round of Hadamard (\cref{prop:S}).}
\label{tab:e4a}
\centering
\small
\begin{tabular}{llcccc}
\toprule
model & input & no rotation & act-order & Hadamard (same basis) & weight-side cost $\Cw$\\
\midrule
Qwen3-1.7B & q/k/v & 0.238 / 0.236 & 0.076 / 0.075 & 0.039 / 0.039 & 0.019\\
 & o\_proj & 0.271 / 0.265 & 0.192 / 0.189 & 0.080 / 0.079 & 0.036\\
 & gate/up & 0.130 / 0.128 & 0.023 / 0.022 & 0.026 / 0.026 & 0.012\\
 & down\_proj & 0.341 / 0.338 & 0.229 / 0.228 & 0.056 / 0.055 & 0.034\\
Llama-3.1-8B-I. & q/k/v & 0.180 / 0.179 & 0.085 / 0.085 & 0.030 / 0.029 & 0.010\\
 & o\_proj & 0.247 / 0.244 & 0.147 / 0.145 & 0.094 / 0.092 & 0.032\\
 & gate/up & 0.030 / 0.030 & 0.012 / 0.011 & 0.018 / 0.018 & 0.005\\
 & down\_proj & 0.137 / 0.136 & 0.065 / 0.065 & 0.024 / 0.024 & 0.006\\
\bottomrule
\end{tabular}
\end{table}

\paragraph{Estimating $\Cw$ from the weights.}
$\Cw$ is estimated from RTN code entropies at a common step on the original and rotated weights, without the Hessian. Over $64$ layers, its deviation from the measured value at $R=4$ has median $0.0008$ bit and maximum $0.034$ bit. The variance part of $\Cw$ (the change in the column-variance AM/GM term) is $0.001$--$0.026$ bit and the negentropy part $0.003$--$0.013$ bit (medians per layer type).

\paragraph{Fixed rate.}
At $b=4$ and $b=3$ with group-128 scales:
\begin{itemize}[leftmargin=*, itemsep=0pt, topsep=1pt]
\item Without rotation, the distortion is higher than with Hadamard by $0.13$--$0.29$ bit-equivalent (median).
\item WaterSIC column multipliers inside the group format change it by $+0.002$ to $+0.008$ bit in the Hadamard basis and by $+0.21$ to $+0.37$ bit in the original basis.
\item Entropy-coded GPTQ with Hadamard reaches the distortion of INT-$b$ with Hadamard at $0.78$--$0.96$ bit below the nominal rate $b+0.125$.
\end{itemize}

\begin{table}[h]
\caption{WikiText-2 perplexity, all decoder linear layers quantized, calibration seed $0$. Rates include side information (scales: $+0.125$; PlainWaterSIC steps: $+0.003$--$0.006$; asymmetric INT zero points are not counted). bf16: Qwen3-1.7B $16.72$, Llama-3.1-8B-Instruct $7.22$.}
\label{tab:ppl-full}
\centering
\small
\begin{tabular}{lccc}
\toprule
configuration & bits & Qwen3-1.7B & Llama-3.1-8B-I.\\
\midrule
entropy-coded GPTQ, no rotation & 2.01 / 2.50 / 3.00 & 152.95 / 30.26 / 23.53 & 41.37 / 30.13 / 12.81\\
entropy-coded GPTQ + Hadamard & 2.01 / 2.50 / 3.00 & 26.93 / 19.60 / 18.56 & 11.18 / 8.55 / 7.85\\
PlainWaterSIC, no rotation & 2.01 / 2.51 / 3.00 & 26.06 / 19.44 / 17.31 & 10.37 / 8.45 / 7.76\\
PlainWaterSIC + Hadamard & 2.51 & 19.69 & 8.61\\
INT3 g128 sym, no rotation / + Hadamard & 3.125 & 62.32 / 23.44 & 21.15 / 9.64\\
INT3 g128 asym, no rotation / + Hadamard & 3.125 & 49.87 / 20.12 & 21.26 / 8.45\\
INT4 g128 sym, no rotation / + Hadamard & 4.125 & 20.36 / 17.45 & 11.97 / 7.52\\
INT4 g128 asym, no rotation / + Hadamard & 4.125 & 19.04 / 17.27 & 8.65 / 7.44\\
\bottomrule
\end{tabular}
\end{table}

\paragraph{Calibration seeds.}
For Qwen3-1.7B with seeds $0/1/2$:
\begin{itemize}[leftmargin=*, itemsep=0pt, topsep=1pt]
\item 3 bits: GPTQ+Hadamard $18.80\pm1.14$, PlainWaterSIC $17.33\pm0.02$, PlainWaterSIC+Hadamard $18.21\pm0.05$ (seeds $1,2$ only);
\item 2.5 bits: GPTQ+Hadamard $20.05\pm0.54$, PlainWaterSIC $19.98\pm1.06$, PlainWaterSIC+Hadamard $20.47\pm1.19$.
\end{itemize}
Rotating PlainWaterSIC is consistently worse, in the direction predicted by $\Cw>0$, but by more than the measured $\Cw$ alone would suggest. Error propagation in the sequential pipeline may contribute; we have not tested this.

\subsection{Column ordering after rotation (E5)}
\label{app:ordering}

For Haar-rotated spectra we compare the mean pivot under the natural order, under act-order (static order by $\diag\Sigma$), and under greedy min-pivot (SQRD~\citep{wubben2001efficient}). On five synthetic laws with $n=256$--$8192$:
\begin{itemize}[leftmargin=*, itemsep=0pt, topsep=1pt]
\item The saving of min-pivot over the natural order fits $C_\nu n^{b}$ with $b\in[-0.52,-0.49]$, with relative residual $\approx5\%$. A $\sqrt{\ln n/n}$ fit has residual $18$--$27\%$.
\item $C_\nu$ ranges from $0.04$ (Marchenko--Pastur) to $5.7$--$6.7$ (two-point).
\item Act-order saves a constant fraction ($1/1.6$ to $1/1.07$) of what min-pivot saves.
\end{itemize}
On the $96$ real spectra at native dimension, min-pivot recovers a median $18\%$ (range $7$--$38\%$) of the rotated gap, i.e.\ $0.002$--$0.043$ bit. Act-order recovers about half of that.

\subsection{Basis gap: lattice reduction and certificates (E6)}
\label{app:lll}

We use Qwen3-1.7B blocks $0$ and $27$, and blocks $0$, middle and last of Qwen3-8B-Base and Llama-3.1-8B-Instruct, with four inputs each. On these layers we computed the following quantities for $\Phi=\sum_k\pi_k$:
\begin{itemize}[leftmargin=*, itemsep=0pt, topsep=1pt]
\item $\Phi$ for the natural order, the min-pivot order, and min-pivot followed by full LLL with $\delta=0.99$ (for $n\le4096$, capped at $600$ s);
\item the AM--GM lower bound $n(\det\Sigma)^{1/n}$;
\item a covering-radius certificate $\Gamma\ge\max_k\sigma_{\min}(A_{II})^2\sum_{j\in I}\pi_j$, where $I$ is the set of the last $k$ Gram--Schmidt coordinates and $A$ the unit-triangular factor.
\end{itemize}
The results:
\begin{itemize}[leftmargin=*, itemsep=0pt, topsep=1pt]
\item \emph{LLL.} After min-pivot ordering, full LLL reduces $\Phi$ by at most $0.37\%$ ($\le0.003$ bit) on all $24$ layers with $n\le4096$. The only reductions above $0.012\%$ occur on o\_proj inputs. The transformation matrices have entries $|T_{ij}|\le7$.
\item \emph{Group-respecting orders.} Orders that keep group-128 columns contiguous increase $\Phi$, and the INT4-g128 loss, by $3$--$53\%$.
\item \emph{Certificate.} The certificate exceeds the AM--GM bound only on Llama block~0's q/k/v input ($1.035\times$). Elsewhere it is $0.03$--$0.39\times$ the bound, and between $0.03$ and $0.31$ of $\Phi_{\text{min-pivot}}$.
\end{itemize}
The remaining gap after ordering is therefore not removable by basis reduction in practice. How it splits into basis and lattice components on real Hessians cannot be decided with this certificate.